# A Generative Model for Extrapolation Prediction in Materials Informatics

*Kan Hatakeyama-Sato\* and Kenichi Oyaizu\**

AUTHOR ADDRESS

Department of Applied Chemistry Waseda University, Tokyo 169-8555, Japan

AUTHOR INFORMATION

**Corresponding Author**

oyaizu@waseda.jp (Kenichi Oyaizu)

ABSTRACT

We report a deep generative model for regression tasks in materials informatics. The model is introduced as a component of a data imputer, and predicts more than 20 diverse experimental properties of organic molecules. The imputer is designed to predict material properties by "imagining" the missing data in the database, enabling the use of incomplete material data. Even removing 60% of the data does not diminish the prediction accuracy in a model task. Moreover, the model excels at extrapolation prediction, where target values of the test data are out of the range of the training data. Such extrapolation has been regarded as an essential technique for exploring novel materials, but has hardly been studied to date due to its difficulty. We demonstrate that the prediction performance can be improved by >30% by using the imputer compared with traditional linear regression and boosting models. The benefit becomes especially pronounced with few records for an experimental property (< 100 cases) when prediction would be difficult by conventional methods. The presented approach can be used to more efficiently explore functional materials and break through previous performance limits.

**1. Introduction**

The aim of materials informatics is to reveal the underlying trends in materials science by using machine learning tools to enable efficient exploration of functional materials, including those for use in energy-related devices.[1-5] Since the properties of materials are uniquely determined by the states of their constituent atoms, their observed structure-property relationships can be mimicked by machine learning models (**Figure 1**).[3-5] Their predictions are often more accurate than traditional theory-based predictions and simulations, especially in the cases of complex systems where the computational costs of the traditional approaches increase exponentially.[5-7]

To describe the structures of materials, researchers typically consider their characteristic representations, such as molecular formulas and crystal structures.[1, 3] In particular, numeric array-type expressions are frequently used in materials informatics because of their high computational processability.[1, 3] Molecular descriptors, fingerprints, and neural network outputs are representative ideas for efficiently expressing the structural information of organic molecules, and could be alternatives to molecular structures or their character strings (e.g., simplified molecular input line entry system: SMILES).[1, 3, 8-11] Machine learning models can connect structure information and material properties via statistical relationships (**Figure 1**).[1] Diverse features, including mechanical properties,[12] permittivity,[6] electric conductivity,[5, 7, 13] thermal conductivity,[14] and photoconversion efficiency[15] have been successfully predicted from structural data with reasonable accuracy.

Although conventional machine learning models can predict properties of materials at higher levels of accuracy than traditional approaches, the processable information has been strictly limited. Usually, one model processes only a single property of a limited number of material species,[1] whereas humans can judge material characteristics not only from the specific species, but also from various material data and general knowledge of science.[5] The lack of such prior knowledge often becomes problematic for prediction models, especially when using small material databases, leading to inferior prediction accuracy. More flexible and human-like data processing algorithms are needed to allow the models to acquire a broad knowledge of materials and achieve more reliable predictions.[5]

Several approaches have been proposed to improve prediction accuracy.[1, 5, 10-11] Transfer learning has become a powerful deep learning technique: here, a neural network is pretrained with a large database, and the model is reused with a smaller target database.[10, 16-17] Increased prediction accuracy has been reported in the fields of image processing,[17] text processing,[16] and even materials science.[7, 10, 18-19] On the other hand, the amount of accessible material data (e.g., $10^1$−$10^4$ cases for one experimental database)[3, 5] might, in general, be too small to be considered as big data for deep learning. In contrast, over $10^7$ records are used in typical transfer learning tasks.[16-17] More efficient approaches are needed to realize robust predictions with smaller databases.

Here, we propose generative models[11, 20-21] for effectively processing broad material data. The models learn the distribution of inputted data, not a specific relationship between an explanatory variable $x$ and target $y$.[20] The introduction of the "imagined" databases made by generative models was found to be a key to predicting various material properties from small experimental material databases, even for databases missing many records. Compared with standard models, generative models led to improved regression accuracy in both interpolation and extrapolation tasks (i.e., for predicted $y$ targets, respectively, within and outside of the range of the training dataset). Except for linear regression, no practical method has been developed to carry out such extrapolations in materials informatics,[1] and hence a promising alternative is proposed in this study. The present results are expected to open a new path for more efficiently exploring new materials that can overcome the performance limits of previous materials.

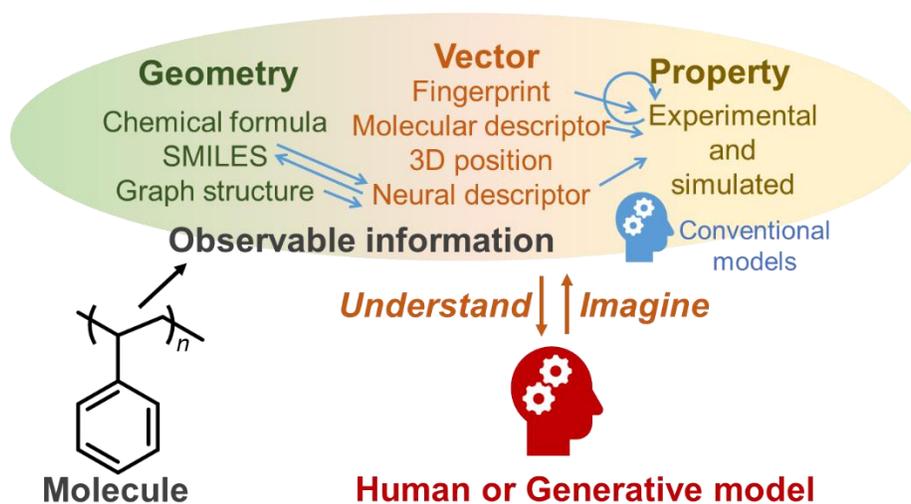

**Figure 1.** Scheme for processing material data. The material information is observable as a chemical formula, experimental properties, and various other characteristics. The conventional machine learning models typically treat the relationship between only two factors (e.g., fingerprint and experimental property). On the other hand, humans and generative models can integrate and imagine versatile information and consider various underlying relationships.

## 2. Descriptor selection

The objective of this study was to introduce a generative model to predict diverse material properties from structural information. Since the model can, in principle, process any numeric information, the ultimate goal was to be able to input any related numeric information on materials (**Figure 1**). On the other hand, due to the benefit of using a smaller input for practical calculations, we screened various descriptors to express material structures as the first step. For organic molecules, widely used numeric inputs are fingerprints,[8] molecular descriptors,[1] output of neural networks (neural descriptors),[5, 9, 11] and simulated structure-related properties (e.g., energy levels of orbitals)[15, 22]. Since these algorithms were designed and developed independently, there seems to be no guidelines for descriptor selection.

Descriptors were screened using a model database of organic compounds. The database contained about 160 types of small molecules consisting of H, C, N, O, S, P, and halogen atoms (**Figure S1**, see Supporting Information for details). Their primary physical properties (boiling point, melting point, density, and viscosity) were recorded as experimental information. Each property was predicted from numeric structural information calculated using various algorithms: two- or three-dimensional geometric molecular descriptors (Desc 2D or 3D),[23] fingerprints (FPs), descriptors considering empirical data (HSPiP),[24] neural descriptors, and single-molecule properties calculated by a semiempirical molecular orbital method (PM7, **Figure S2**, see Supporting Information for details). For neural descriptors, we pretrained neural networks to predict four parameters from molecular structures expressed as graph structures (i.e., transfer learning, **Figure S3**). All variables were scaled to a range of 0 to 1 for normalization.

After randomly splitting the database into the training (80%) and test (20%) datasets, an extreme gradient boosting (XGB)[25] regressor was trained to predict each target property from the generated descriptors (**Figure S4**). XGB is considered one of the most promising algorithms for adequately treating nonlinear interactions, even with small databases.[25] As the control to using only the descriptors, the other three experimental properties were added as explanatory variables (e.g., prediction of the boiling point from the molecular descriptors melting point, density, and viscosity).

Mean absolute errors (MAEs) of the actual and predicted values for the test dataset are summarized in **Figure 2a** and **S5**. For boiling point, the HSPiP descriptor offered the best prediction performance. Neural descriptors pretrained for boiling point were also useful for predicting the parameter. We anticipated that the neural descriptor made for melting point would also be beneficial for predicting boiling point because the two parameters correlate strongly. However, the loss did not improve significantly compared with the viscosity descriptors. The results indicate that transfer learning in materials science is not always promising, certainly because of the wide variety of predictable parameters and small experimental databases available for pretraining. In contrast, the prediction accuracy improved when the experimental parameters were included in $x$, where models can consider the relationships between properties more directly (**Figure 2a**, blue bars).

When the target parameter was changed from boiling point, the useful descriptors for the prediction changed dramatically (**Figure S5**). In other words, no all-purpose descriptors were found even for this simple task. One universal trend we detected was that the inclusion of experimental parameters in $x$ reduced (or at least did not affect) the prediction errors, indicating that the experimental properties are highly useful descriptors.

To reveal the most influential factors in the prediction accuracy, we calculated the absolute correlation coefficients between the target values and descriptors (**Figure 2b** and **S6**).

The obtained box plots indicate that better prediction performance was more easily obtained with the descriptors providing the higher coefficients against *y* (e.g., some coefficients for HSPiP exceeded 0.6, and were higher than those provided by any of the other descriptors). For a more precise comparison, regressions were repeated using five explanatory parameters randomly selected from all of the descriptors (**Figures S7** and **S8**). There was an apparent relationship between the averages of the five coefficients and the MAE. This result meant that the predictions were mostly influenced by the linearity between ***x*** and *y*, and not by more complicated trends such as nonlinear interactions. The same results were observed even in the extrapolation tasks, where the records of the top 20% of *y* values were used as test datasets (**Figure S8**; a linear model was used for the reason discussed in the next section). For instance, no MAE below 0.01 was indicated unless the average coefficient exceeded about 0.2 during the boiling point prediction (**Figure S8a**). In summary, we found that the availability of explanatory variables that correlated strongly with *y* was essential for the machine learning, especially with small databases, and that the availability of experimentally derived properties of materials was useful for satisfying such requirements.

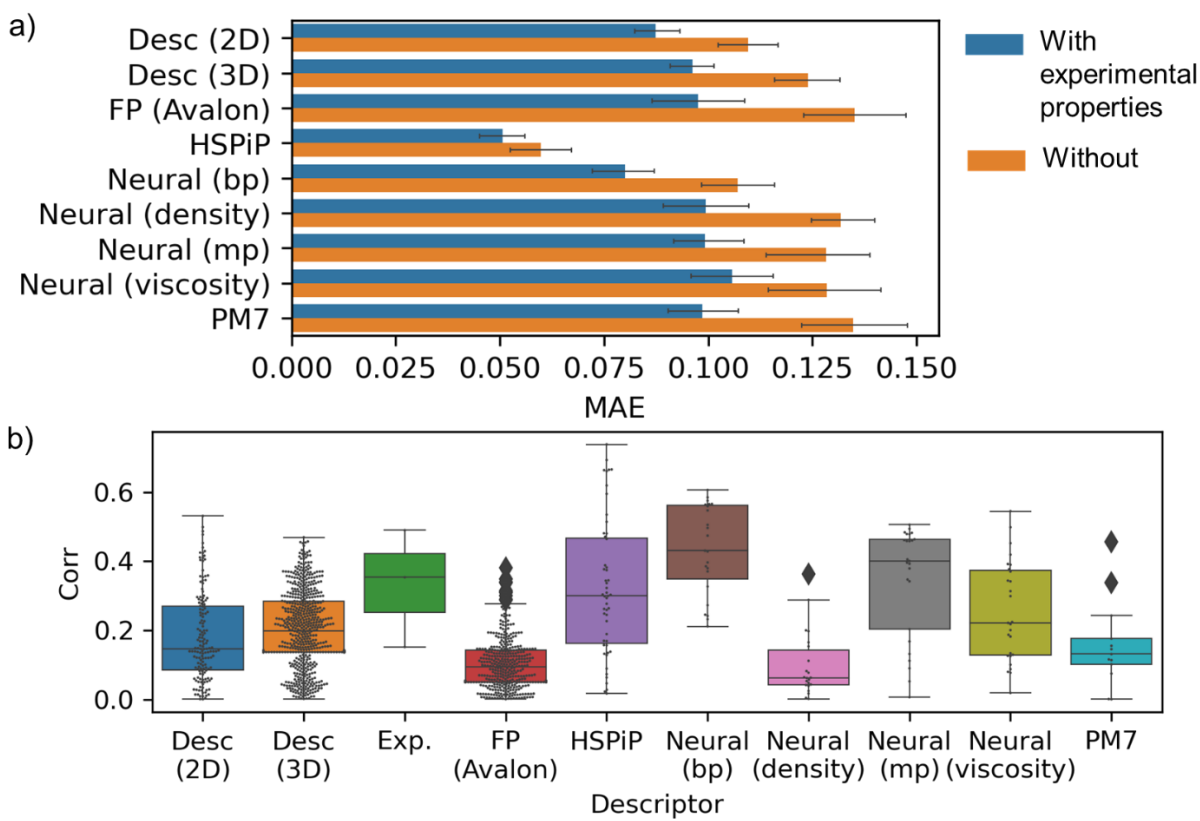

**Figure 2.** Prediction results of boiling point using various descriptors. XGB regressor was used for prediction. a) MAE for the test datasets. Melting point, density, and viscosity were included in $x$ with blue bars, whereas only the descriptors were used with the orange bars. Results for the 5-hold cross-validation are shown. Error bars indicate standard errors. b) Distribution of absolute correlation coefficients between the explanatory variables and boiling point. Exp, bp, and mp mean experimental properties, boiling point, and melting point. A detailed explanation of the descriptors is shown in the Supporting Information.

### 3. Introduction of a generative model for regression

Although experimental properties can be beneficial explanatory variables, the cost of measuring them is typically much higher than the costs of standard molecular descriptors and simulations.[1, 7] For this reason, experimental databases tend to have many missing values.[5] However, regular machine learning models cannot treat them appropriately.[25-26]

Generative models are attracting attention because of their robustness against missing data.[27-28] For instance, they can repair worm-eaten images of handwritten numbers (**Figure 3a**).[27-28] Even if a generative model is trained with only worm-eaten images, the model can estimate ("imagine") the original images by considering the relationships among the observed variables, as humans do.[27-28] We anticipated that the same approach could also be used with materials, that is to estimate unobserved (missing) parameters from observed material descriptors and properties (**Figure 1** and **Figure 3a**).

Regression by generative models was carried out by so-called imputation processes (**Figure 3b**).[27-28] First, all missing data, including target variable $y$, were replaced with constant values. Second, a generative model $f_{ML}$ was trained with the records to mimic the data distribution ($z_{old}$). Then, the initially missing parameters were replaced with the new predicted value $z_{new} = f_{ML}(z_{old})$. The learning and replacement were repeated so that the updated $z$ reached convergence. After imputation, a complete database was constructed with the most feasible (or imagined) values predicted by the generative model. This process is more human-like than conventional regressions because well-experienced researchers often predict properties of materials from incomplete material data by "imagining" missing information based on their experiences.

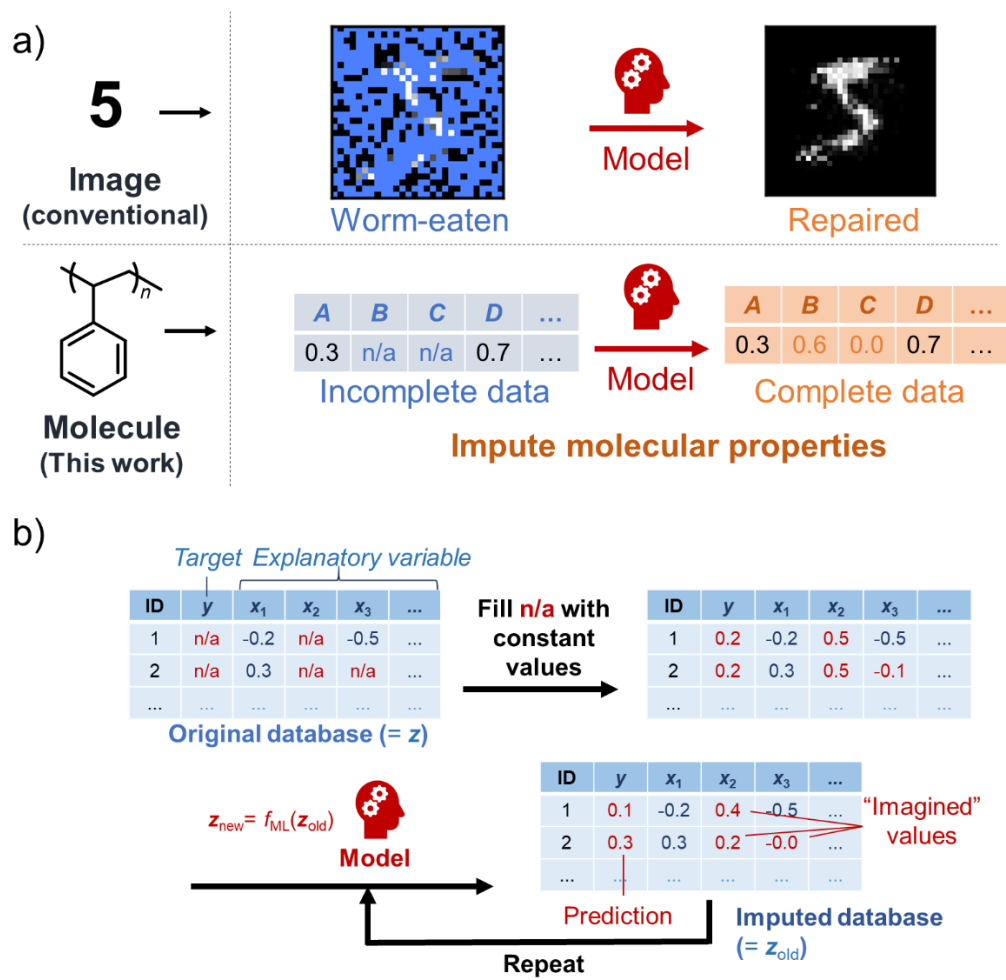

**Figure 3.** Concept of imputation. a) Worm-eaten images and incomplete material data can be repaired using imputers. b) Imputation process. The imputer does not distinguish missing values in $x$ and target value $y$.

By using the database described in the previous section, regression models were constructed to predict the four parameters from the structural information. Here, two-dimensional geometric molecular descriptors (Desc 2D) were selected as $x$. We deleted 0%, 30%, or 60% of the explanatory variables randomly to introduce missing values. Then, the database was split into the training and test datasets in order: the records with the top 20% of target values were used for the test and the rest were used for the training (i.e., for the extrapolation task).

A recently reported framework of a flow-based model[21] called Monte Carlo flow (MCFlow)[28] was introduced as the imputer. The framework has achieved state-of-the-art performance for a series of imputation tasks, including image restoration.[28] The adequate randomness and reversible mapping functions in the framework played essential roles in the prediction.[28] Compared with other representative deep generative models, including autoencoders and the generative adversarial network (GAN), this framework achieved more accurate imputation with smaller databases.[27-28]

For missing ratios (i.e., proportions of the data missing) of up to at least 0.6, the MAE of the predicted boiling point for the test dataset was less than 0.2 (**Figure 4**). In contrast, XGB, which can process missing values directly,[25] yielded a much larger MAE of about 0.3 when 60% of the explanatory data were deleted. We note that the $y$ predicted by XGB was always in the range of the training dataset, meaning that the predictor was useless for the extrapolation task. Most decision-tree-type regressors, including random forest and other boosting models,[26] could not be used for the extrapolation tasks because of their unique classification-based algorithms (and showed results similar to those obtained when using XGB, data not shown).[25]

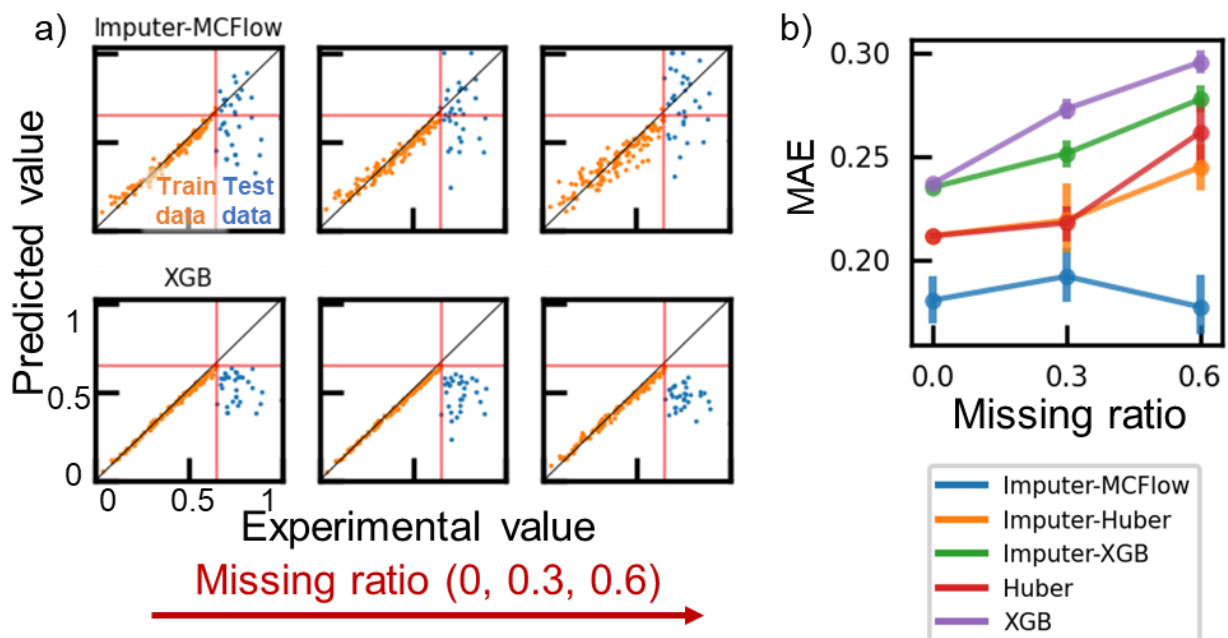

**Figure 4.** Boiling point prediction by MCFlow and XGB regressors. a) Experimental and predicted values. 0%, 30%, or 60% of the explanatory variables were filled as missing values randomly. b) MAE for the test datasets with various regression conditions. Mean values after different 5-times data preparation and regression are shown. Error bars indicate standard errors. Selected results for an extrapolating task are shown in this Figure. Full results are shown in Figures **S9** and **S10**.

Instead of XGB, a simple linear regression model was examined for extrapolation. After replacing missing data with mean values, regression models were made using the Huber loss,[26, 29] which is robust to outliers (**Figures S9** and **S10**). Here, MAEs of the predicted values were found to range from around 0.20 to 0.25, better than the results when using XGB but worse than the results when using MCFlow. To our knowledge, linear regression and related algorithms have been the only practical approach for extrapolation tasks in material science.[1, 10, 30] However, the deep-flow-based model has now become a promising alternative, one also excelling at solving the missing data issue.

The prediction process in MCFlow was analyzed by visualizing the internal variables in the model (**Figure S11**). The generative model first used the mapping function $f_{map}$ to convert the original input data $z$ to a same-dimensional vector $\mathbf{z_{map}}$ (**Figure S11a**). For regularization, the internal variables are supposed to provide a multivariate normal distribution in flow-based models.[21, 28] Then, a multilayer perceptron was used to convert $\mathbf{z_{map}}$ to $\mathbf{\hat{z}_{map}}$ for the largest possible density estimation. Then, the inverse function $f_{map}^{-1}$ was used to reconstruct $z$. (Note that $f_{map}$ is designed to be invertible in flow-based models.) Interestingly, the internal variables tended to correlate with $y$ more strongly after the conversion (**Figures S11c and d**). This capability of extracting the underlying linearity between the explanatory variables and the target value was considered a key to achieving successful extrapolation.

As a control for MCFlow, we also examined various imputation algorithms for regression. Other representative generative models were examined, namely, GAN,[27] an autoencoder (AE),[20] and a variational autoencoder (VAE)[7, 11] (**Figure S9e**). The prediction using GAN did not vary with changing $x$, meaning that the training was unsuccessful; the deep learning model typically requires over $10^4$ cases of training data,[27] whereas only about 150 cases were given in this task. Regular autoencoders (AE and VAE) displayed moderate prediction accuracy for the training dataset, but were unable to make predictions in the extrapolation regions of the test dataset. The significant difference between the flow-based models and the autoencoders is related to whether or not the mapping function $f_{map}$ is invertible.[21, 28] During machine learning, the invertibility (i.e., regularization) should have positively worked to extract the linearity, but the regular autoencoders suffered from overfitting to the training data in the interpolation region. More detailed processes for successful extrapolation prediction will be studied in future research.

Regular regression models such as XGB can also be used as components in imputers. Imputation was done by preparing individual regression models for each $z_i$ and repeating the regression (**Figure 3**). MissForest, a random forest imputation framework, provides good prediction performance even with small databases.[31] The framework could also be considered a derivative of generative models, because they can estimate the distribution of $z$ from inputted data. Here, XGB and Huber regressors were selected as predictor components. When the missing data of the explanatory variables were filled in, the MAE of the predicted $y$ was generally improved compared with the normal approach of substituting mean values for missing data (**Figure 4b**, **S9**, and **S10**).

Imputers provided improved performance even for the other prediction tasks. Smaller MAEs were obtained with the imputers than when using the regular Huber regression with the other target values (e.g., MAEs of 0.23 and 0.34 when using, respectively, MCFlow and Huber for melting point prediction with 60% data loss, **Figures S9** and **S10**). In the cases of the interpolation tasks, the best results were obtained when XGB imputers were used (**Figures S9** and **S10**) owing to the accurate prediction by XGB in the interpolation region. We believe that carrying out an optimal tuning of MCFlow (e.g., changing hyperparameters and network structures) would improve its prediction accuracy to be as high as XGB.

In summary, for the generative model used for regression, we introduced the imputers to predict material properties directly. The advantage of "imagining" missing data by using generative models was essential for improving the regression performance for the material data.

**4. Predicting multiple properties from an integrated material database**

To demonstrate the practical benefit of generative models, we used various experimental databases to predict multiple properties of materials. In previous work, we collected and integrated various experimental material databases of structure-property relationships of organic molecules and polymers.[5] An updated database was made in this study, covering about 12000 compounds and over 25 types of experimental properties, such as glass transition temperature, heat capacity, vapor pressure, and viscosity (**Figure S1b**). Since the original data sources focused on different material properties, the integrated database included a substantial amount of missing data (**Figure 5a**). However, we have already shown that even incomplete databases of material properties can be useful after appropriate imputation.

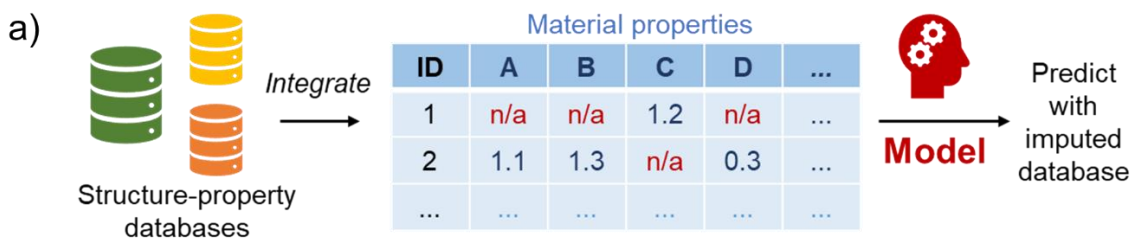

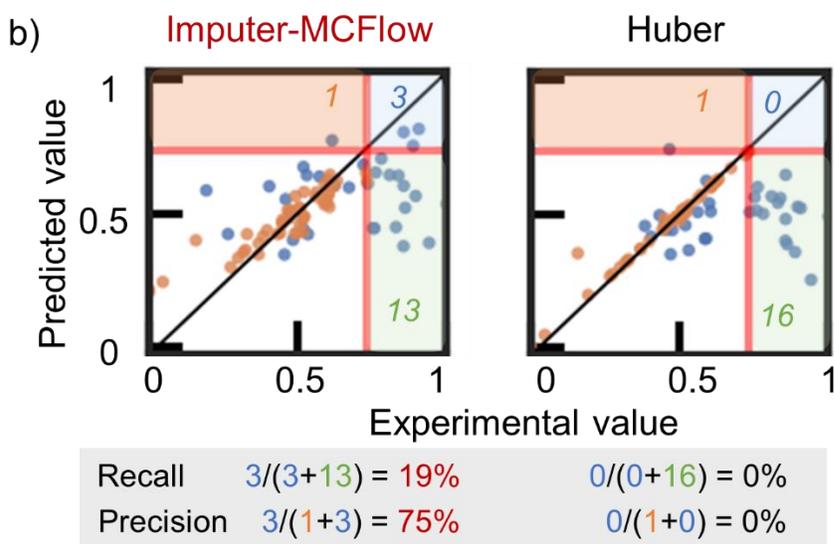

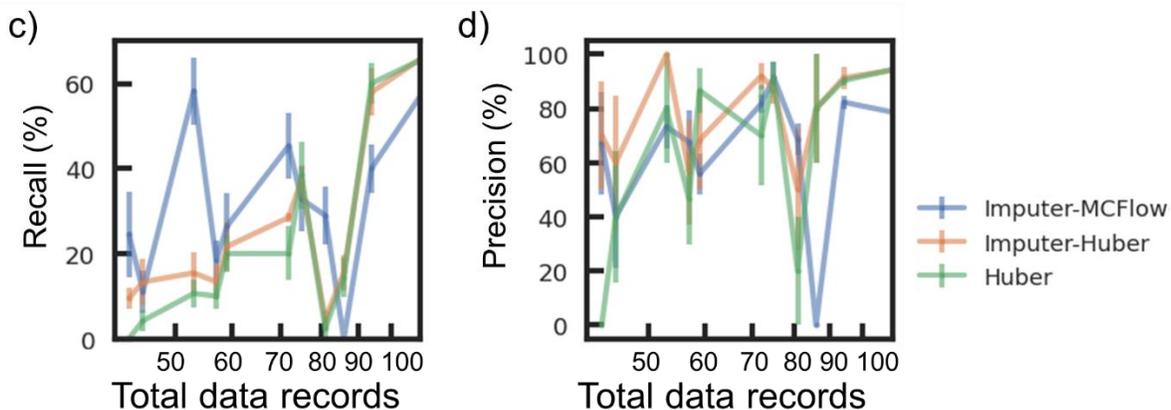

**Figure 5.** Prediction results for an integrated database. a) Scheme to fill missing values in the integrated database. b) Viscosity prediction by MCFlow and Huber regressors. Orange and blue plots represent the train and test datasets. c), d) Recall and precision to extract the extrapolating region materials. Full results are shown in Figures **S13** and **S14**.

Regression tasks were repeated to predict each property from the 2D molecular descriptors and the rest of the material properties (**Figure 5b** and **S12**). Here, the top 20% of target variables and a randomly selected 20% of the records were used as test datasets to evaluate both extrapolation and interpolation prediction performance. In a typical result, viscosity was predicted using about 100 available records (**Figure 5b**). MCFlow achieved higher prediction accuracy than Huber in the extrapolation region (MAEs of about 0.2 and 0.3, respectively, corresponding to >30% improvement), whereas similar results were obtained in the interpolation region.

Regarding other practical criteria for the extrapolation prediction, we assessed recall and precision, which are often used for classification tasks.[26] Recall (resp., precision) was defined as the percentage of successfully predicted cases in the extrapolation region of the experimental (resp., predicted) values (**Figure 5b**). In exploration of novel materials, a higher recall means that novel candidates can be detected with greater accuracy, and a higher precision is needed to avoid improperly detecting candidates whose actual $y$ exists within the range of training values. MCFlow provided a recall of 19% and precision of 75% for density prediction, whereas Huber gave both precision and recall of zero: in other words, the generative model achieved successful extrapolation that could not be done by a conventional approach. We note that realizing both 100% recall and precision is not feasible because no one can perfectly judge whether a new material exceeds the performance limitations of conventional species. Considering the difficulty of the task, the recall of 19% by MCFlow should be considered sufficient as a first step.

The use of imputers improved prediction accuracy even with other target parameters (**Figure 5c**, **5d**, **S13**, and **S14**). As a general trend, MCFlow exhibited a higher recall than Huber for properties with fewer than 100 records (> 20% recall by MCFlow in most cases, **Figure 5c**).

Apart from MCFlow, the Huber imputer also increased the recall and precision with few records, indicating that "imagining" the missing data was essential for the prediction (**Figure 5c** and **d**). The prediction performance of MCFlow and Huber was similar when the number of available records was greater than 100, the reason being that sufficient information was given for the predictors (**Figure S14**). Since experimental databases are often small due to the high cost of collecting data, the newly introduced approach to integrate databases and impute missing data helps researchers explore materials more efficiently.

## 5. Comparison with conventional prediction models in materials informatics

Finally, we compared our new approach to others used in materials informatics (**Figure 6**). The lack of sufficient data for materials is a critical obstacle to realizing effective predictions of their properties and thus requires appropriate machine learning strategies. One promising approach involves a simplification, specifically the use of linear regressions and sparse modeling to robustly treat statistical relationships among materials,[1, 30] which was also demonstrated by Huber regression in this study. However, not every phenomenon can be explained based on simply an assumption of linearity because material properties appear to result from highly complex molecular interactions.[7] More complex models are needed to compensate for the limitations of the linearity-based approach.

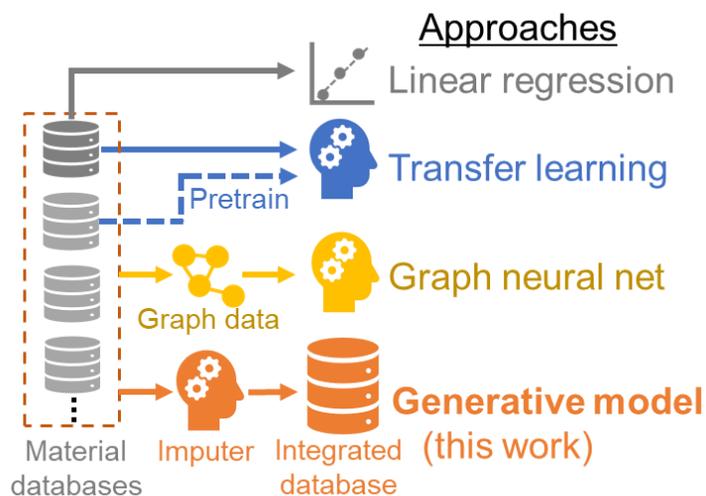

**Figure 6.** Comparison of generative models with other approaches for material property prediction tasks.

Deep learning that includes a generative model as in this study is the opposite idea of sparse modeling. Few- or even zero-shot learning has been achieved by transfer learning of big data.[16-17] Still, transfer learning may not be effective in all cases in materials science, because of the lack of data records as discussed in previous sections. Since a general-purpose transfer learnable material database is not available, researchers must carefully select a proper pretraining database from a massive number of candidates.[7, 10, 19] In essence, this problem becomes the difficult task of hyperparameter selection from an extensive search optimization space.

As a more general approach, the use of more universal graph databases has also been proposed.[5, 32-33] We have demonstrated that diverse material data can be described in universal-format graphs and processed using a single model.[5] The trained model can implicitly recognize the hidden relationships among the structural and property data from different databases. If a sufficiently large amount of data and a powerful learning framework are given, the trained model would be able to respond to any material-related questions accurately.[5] However, the actual lack of such resources still limits the potential of the approach. Furthermore, the black-box prediction prevents researchers from understanding the trends of the structure-property relationships.[5]

The generative model proposed here takes advantage of both transfer learning and graph models. Here, the structure-property relationship can be processed explicitly as "imagined databases." For accurate prediction with small databases, directly considering property-property relationships from the imagined databases was favored over indirectly correlating them by transfer learning or a graph model (e.g., simple linear relationships were essential for regression, **Figure 2**). Furthermore, the flow-based generative model showed the promise of extrapolation predictions, whereas most studies and algorithms of materials informatics have focused on only interpolation tasks.[14]

## 6. Conclusion

We introduced generative models for performing regression tasks in materials informatics. The models could impute and predict diverse information about materials. Prediction accuracy was improved by "imagining" the missing values of the data, which is believed to be a process similar to the one in humans. The approach was beneficial for the extrapolation prediction tasks even with small databases. This approach gave better prediction performance than that observed for traditional linear regression (e.g., >30% improvement of prediction loss), whereas linear models have been only one practical strategy for extrapolation. Extrapolation is critical to find new candidates that break through the performance limits of known materials. The presented approach is expected to open a path to make explicit connections to the knowledge of diverse materials, which is necessary to develop understandable machine learning tools and to realize more general prediction models for materials science.

NOTES

The authors declare no competing financial interest.

ACKNOWLEDGMENT

This work was partially supported by Grants-in-Aid for Scientific Research (Nos. 17H03072, 18K19120, 18H05515, and 19K15638) from MEXT, Japan. The work was also partially supported by the Research Institute for Science and Engineering in Waseda University, and a research grant from the Center for Data Science in Waseda University, and Information Services International-Dentsu, Ltd.

REFERENCES

(1) Ramprasad, R.; Batra, R.; Pilania, G.; Mannodi-Kanakkithodi, A.; Kim, C., Machine learning in materials informatics: recent applications and prospects. *Npj Comput. Mater.* **2017,** *3*, 54.
(2) Sanchez-Lengeling, B.; Aspuru-Guzik, A., Inverse molecular design using machine learning: Generative models for matter engineering. *Science* **2018,** *361*, 360-365.
(3) de Pablo, J. J.; Jackson, N. E.; Webb, M. A.; Chen, L.-Q.; Moore, J. E.; Morgan, D.; Jacobs, R.; Pollock, T.; Schlom, D. G.; Toberer, E. S.; Analytis, J.; Dabo, I.; DeLongchamp, D. M.; Fiete, G. A.; Grason, G. M.; Hautier, G.; Mo, Y.; Rajan, K.; Reed, E. J.; Rodriguez, E.; Stevanovic, V.; Suntivich, J.; Thornton, K.; Zhao, J.-C., New frontiers for the materials genome initiative. *Npj Comput. Mater.* **2019,** *5*, 41.
(4) Noh, J.; Gu, G. H.; Kim, S.; Jung, Y., Machine-enabled inverse design of inorganic solid materials: promises and challenges. *Chem. Sci.* **2020,** *11*, 4871-4881.
(5) Hatakeyama-Sato, K.; Oyaizu, K., Integrating multiple materials science projects in a single neural network. *Commun. Mater.* **2020,** *1*, article number: 49.
(6) Mannodi-Kanakkithodi, A.; Chandrasekaran, A.; Kim, C.; Huan, T. D.; Pilania, G.; Botu, V.; Ramprasad, R., Scoping the polymer genome: A roadmap for rational polymer dielectrics design and beyond. *Mater. Today* **2018,** *21*, 785-796.

(7) Hatakeyama-Sato, K.; Tezuka, T.; Umeki, M.; Oyaizu, K., AI-Assisted Exploration of Superionic Glass-Type Li(+) Conductors with Aromatic Structures. *J. Am. Chem. Soc.* **2020,** *142*, 3301-3305.
(8) Rogers, D.; Hahn, M., Extended-connectivity fingerprints. *J. Chem. Inf. Model.* **2010,** *50*, 742-754.
(9) Zhou, J.; Cui, G.; Zhang, Z.; Yang, C.; Liu, Z.; Wang, L.; Li, C.; Sun, M., Graph Neural Networks: A Review of Methods and Applications. **2018,** *arXiv:1812.08434*.
(10) Yamada, H.; Liu, C.; Wu, S.; Koyama, Y.; Ju, S.; Shiomi, J.; Morikawa, J.; Yoshida, R., Predicting Materials Properties with Little Data Using Shotgun Transfer Learning. *ACS Cent. Sci.* **2019,** *5*, 1717-1730.
(11) Gomez-Bombarelli, R.; Wei, J. N.; Duvenaud, D.; Hernandez-Lobato, J. M.; Sanchez-Lengeling, B.; Sheberla, D.; Aguilera-Iparraguirre, J.; Hirzel, T. D.; Adams, R. P.; Aspuru-Guzik, A., Automatic Chemical Design Using a Data-Driven Continuous Representation of Molecules. *ACS Cent. Sci.* **2018,** *4*, 268-276.
(12) Chen, C.-T.; Gu, G. X., Machine learning for composite materials. *MRS Commun.* **2019,** *9*, 556-566.
(13) Matsubara, M.; Suzumura, A.; Ohba, N.; Asahi, R., Identifying superionic conductors by materials informatics and high-throughput synthesis. *Commun. Mater.* **2020,** *1*, article number: 5
(14) Wu, S.; Kondo, Y.; Kakimoto, M.-a.; Yang, B.; Yamada, H.; Kuwajima, I.; Lambard, G.; Hongo, K.; Xu, Y.; Shiomi, J.; Schick, C.; Morikawa, J.; Yoshida, R., Machine-learning-assisted discovery of polymers with high thermal conductivity using a molecular design algorithm. *Npj Comput. Mater.* **2019,** *5*, 66.
(15) Nagasawa, S.; Al-Naamani, E.; Saeki, A., Computer-Aided Screening of Conjugated Polymers for Organic Solar Cell: Classification by Random Forest. *J. Phys. Chem. Lett.* **2018,** *9*, 2639-2646.
(16) Brown, T.; Mann, B.; Ryder, N.; Subbiah, M.; Kaplan, J.; Dhariwal, P.; Neelakantan, A.; Shyam, P.; Sastry, G.; Askell, A.; Agarwal, S.; Herbert-Voss, A.; Krueger, G.; Henighan, T.; Child, R.; Ramesh, A.; Ziegler, D.; Wu, J.; Winter, C.; Hesse, C.; Chen, M.; Sigler, E.; Litwin, M.; Gray, S.; Chess, B.; Clark, J.; Berner, C.; McCandlish, S.; Radford, A.; Sutskever, I.; Amodei, D., Language Models are Few-Shot Learners. **2020,** *arXiv:2005.14165*.
(17) Dosovitskiy, A.; Beyer, L.; Kolesnikov, A.; Weissenborn, D.; Zhai, X.; Unterthiner, T.; Dehghani, M.; Minderer, M.; Heigold, G.; Gelly, S.; Uszkoreit, J.; Houlsby, N., An Image is Worth 16x16 Words: Transformers for Image Recognition at Scale. **2020,** *arXiv:2010.11929*.
(18) Hutchinson, M.; Antono, E.; Gibbons, B.; Paradiso, S.; Ling, J.; Meredig, B., Overcoming data scarcity with transfer learning. **2017,** *arXiv:1711.05099*.
(19) Lee, J.; Asahi, R., Transfer learning for materials informatics using crystal graph convolutional neural network. **2020,** *arXiv:2007.09932*.
(20) Kingma, D.; Rezende, D.; Mohamed, S.; Welling, M., Semi-supervised Learning with Deep Generative Models. **2014,** *arXiv:1406.5298*.
(21) Ho, J.; Chen, X.; Srinivas, A.; Duan, Y.; Abbeel, P., Flow++: Improving Flow-Based Generative Models with Variational Dequantization and Architecture Design. **2019,** *arXiv:1902.00275*.
(22) Jinich, A.; Sanchez-Lengeling, B.; Ren, H.; Harman, R.; Aspuru-Guzik, A., A Mixed Quantum Chemistry/Machine Learning Approach for the Fast and Accurate Prediction of Biochemical Redox Potentials and Its Large-Scale Application to 315000 Redox Reactions. *ACS Cent. Sci.* **2019,** *5*, 1199-1210.

(23)	RDKit: Open-source cheminformatics; http://www.rdkit.org
(24)	HSPip: Hansen Solubility Parameters in Practice; https://www.hansen-solubility.com/HSPiP/
(25)	Chen, T.; Guestrin, C., XGBoost: A Scalable Tree Boosting System. In *Proceedings of the 22nd ACM SIGKDD International Conference on Knowledge Discovery and Data Mining*, 2016; pp 785-794.
(26)	Pedregosa, F.; Varoquaux, G.; Gramfort, A.; Michel, V.; Thirion, B.; Grisel, O.; Blondel, M.; Prettenhofer, P.; Weiss, R.; Dubourg, V.; Vanderplas, J.; Passos, A.; Cournapeau, D.; Brucher, M.; Perrot, M.; Duchesnay, E., Scikit-learn: Machine Learning in Python. *J. Mach. Learn. Res.* **2011,** *12*, 2825-2830.
(27)	Yoon, J.; Jordon, J.; Schaar, M.; Xu, B.; Bernal, E., GAIN: Missing Data Imputation using Generative Adversarial Nets. **2018,** *arXiv:1806.02920*.
(28)	Richardson, T.; Wu, W.; Lin, L.; Xu, B.; Bernal, E., MCFlow: Monte Carlo Flow Models for Data Imputation. **2020,** *arXiv:2003.12628*.
(29)	Huber, P. J., Robust Estimation of a Location Parameter. *Ann. Math. Stat.* **1964,** *35*, 73-101.
(30)	Nakada, G.; Igarashi, Y.; Imai, H.; Oaki, Y., Materials‐Informatics‐Assisted High‐Yield Synthesis of 2D Nanomaterials through Exfoliation. *Adv. Theory Simul.* **2019,** *2*, 1800180.
(31)	Stekhoven, D. J.; Buhlmann, P., MissForest--non-parametric missing value imputation for mixed-type data. *Bioinformatics* **2012,** *28*, 112-8.
(32)	Mrdjenovich, D.; Horton, M. K.; Montoya, J. H.; Legaspi, C. M.; Dwaraknath, S.; Tshitoyan, V.; Jain, A.; Persson, K. A., propnet: A Knowledge Graph for Materials Science. *Matter* **2020,** *2*, 464-480.
(33)	Schwaller, P.; Petraglia, R.; Zullo, V.; Nair, V. H.; Haeuselmann, R. A.; Pisoni, R.; Bekas, C.; Iuliano, A.; Laino, T., Predicting retrosynthetic pathways using transformer-based models and a hyper-graph exploration strategy. *Chem. Sci.* **2020,** *11*, 3316-3325.

# A Generative Model for Extrapolation Prediction in Materials Informatics

*Kan Hatakeyama-Sato\* and Kenichi Oyaizu\**

Department of Applied Chemistry Waseda University, Tokyo 169-8555, Japan

\*oyaizu@waseda.jp

**Experimental methods**

**Data availability**

Databases and Python codes used in this study can be accessed at a repository (https://github.com/KanHatakeyama/gen_model). All programs were made with Python unless noted otherwise.

**Computer**

A desktop computer (Intel Core i9-9900K CPU @ 3.60 GHz, 32 GB memory, GeForce RTX2080 graphical processing unit) was used for data processing and machine learning.

**Database preparation**

A small compound database was constructed based on our previous paper, which integrated open material data from public literature.[1] The database contains experimental physical properties (boiling temperature, melting temperature, density, and viscosity) of about 160 small chemicals consisting of H, C, N, O, S, P, and halogen atoms. Chemical structures were recorded by simplified molecular input line entry system (SMILES). An integrated database was prepared similarly. Properties with over 40 records were used for machine learning. Further information is described in the database file. A summary of the databases is shown in **Figure S1**. Chemical structures of repeating units were considered as molecular structures for polymers.

**Data processing**

Before machine learning, all numeric variables were normalized to the range of [0,1] by Min-Max scalers. For regression, a series of molecular descriptors were calculated as $x$ (**Figure S2**). The relevant program codes are available at the repository.

 a) FP(Avalon)

  512-Dimensional fingerprint calculated by an Avalon algorithm using an open-source python package of RDKit (version 2018.09.1, https://www.rdkit.org/).

 b) Desc(2D)

Basic 200-dimensional continuous molecular descriptors, implemented in a "Descriptors.descList" class of RDKit. The descriptors did not consider three-dimensional molecular structures.

c) Desc(3D)

629-Dimensional continuous molecular descriptors, implemented in "CalcAUTOCORR3D", "CalcMORSE," "CalcRDF," and "CalcWHIM" classes of RDKit. The descriptors were calculated according to the three-dimensional structures of the molecules.

d) HSPiP

56-Dimensional continuous molecular descriptors calculated by a Hansen Solubility Parameters in Practice package (HSPiP version 5.3.05, https://www.hansen-solubility.com/HSPiP/).

e) PM7

13-Types of molecular properties (energy levels, dipole moments, heat formation, and magnetic moment) calculated by a semi-empirical calculation using the PM7 Hamiltonian.

f) Neural

32- Dimensional continuous numeric arrays calculated by graph neural networks, which were trained to predict boiling temperature, melting temperature, density, or viscosity of all compounds recorded in the small database (**Figure S3**). Properties were predicted from chemical structures represented by graphs. We note that data leakages occurred when predicting a target value from the same-type neural descriptors during the main regression tasks (e.g., predict boiling temperature from Neural(bp) in **Figure S5**), which somewhat unfairly improved the prediction performances compared to the other descriptors. In practical cases, pretraining must be done only with train datasets, whereas this study only focused on the effect of transfer learning of different target variables.

**Regression models**

Following regression models were employed to predict the properties of molecules. The relevant program codes are available at the repository.

a) XGB (eXtreme Gradient Boosting)

Open Python library of XGB (version 0.90, https://xgboost.readthedocs.io/) was used. Regressions were conducted with default hyperparameters. Missing values were used as they were.

b) Huber

A Huber regression class in scikit-learn (version 0.23.2, https://scikit-learn.org/) was used. Regressions were conducted with default hyperparameters. Before machine learning, missing values were filled with mean values for each $x_i$.

c) Imputer-Huber/-XGB

A modified module of MissForest[2] was prepared. Instead of using RandomForest, Huber or XGB regressors were used as prediction models. Each column was imputed in parallel to reduce calculation time, while the previous one calculated columns in sequence.

d) Imputer-MCFlow

A previously reported MCFlow program[3] was modified so that it could be used as an imputer class facility. The core algorithms were unchanged.

e) Other imputers

As a generative adversarial network-based imputer, a reported framework of generative adversarial imputation nets (GAIN)[4] was employed. Autoencoder (AE) and variational autoencoder (VAE) were constructed by connecting an input layer, a 10-dimensional hidden layer, and an output layer.

**Regression task 1: Comparison of descriptors (Figure S5)**

The four types of experimental molecular properties (boiling temperature, melting temperature, density, and viscosity) recorded in the small compound database were predicted from different descriptors shown in **Figure S2**. XGB regressor was used for prediction. The database was split into the train (80%) and test (20%) datasets randomly. Only the train dataset was used to train the model. The rest of the three molecular properties were also included in the explanatory variables for comparison (**Figure S5**, blue bars). 5-Hold cross-validation was conducted to analyze the statistical trends. We used mean absolute error (MAE) between the predicted and exact values to evaluate the prediction performance.

**Regression task 2: Random selection of descriptors (Figure S8)**

Similar to task 1, experimental molecular properties were predicted from descriptors. In this case, only five parameters were selected from all descriptors and used as explanatory variables (**Figure S7**). XGB regressor was used for prediction, and MAE for a test dataset was evaluated. The random parameter selection and prediction processes were repeated many times. To analyze the relationship between the "quality" of the explanatory parameters and the prediction performance, a new parameter was defined: the average value of the absolute correlation coefficients between the target value ($y$) and the explanatory variable ($x_i$).

**Regression task 3: Prediction with missing values (Figure S9)**

The four physical properties were predicted from Desc(2D) descriptors. For the interpolation task, the small compound database was split into the train (80%) and test (20%) datasets randomly. In contrast, records with the top 20% target values were used as a test and the rest as a train for the extrapolation tasks. 0, 30, or 60% of the explanatory variables were replaced with missing values (n/a) randomly. The data split and prediction processes were repeated 5 times. XGB and Huber repressors were used as predictors for interpolation and extrapolation tasks, respectively.

**Regression task 4: Prediction with the integrated database (Figure S12)**

The properties of molecules recorded in the integrated database were predicted from "Desc(2D)" descriptors. Experimental properties except for the target value were also included in $x$. The missing experimental values were treated according to the procedures described in the regression models section. For interpolation tasks, the database was split into the train (80%) and test (20%) datasets randomly. For extrapolation tasks, records with the top 20% values were extracted as the test. Further, 20% of the data were selected randomly in the interpolation region. Thus, the overall train and test ratio was 60/40. The data splitting and regressions were repeated 5 times. Properties containing over 1000 records were not predicted by Imputer-Huber because of the high calculation cost (e.g., the calculation will take more than 100 hours with 10000 records, whereas MCFlow took less than one hour).

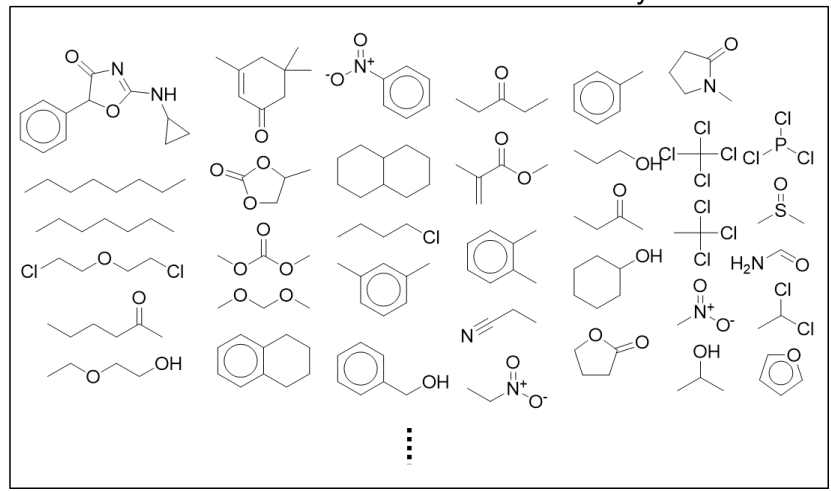

a)

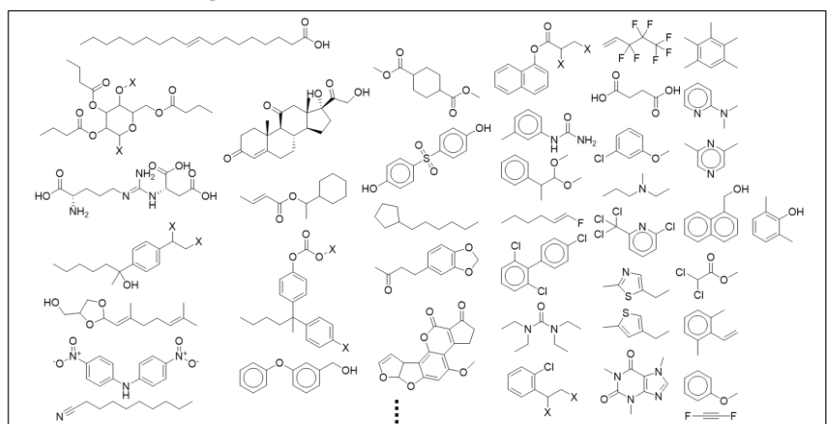

b)

**Figure S1.** Information of chemical compound database used in this study. a) Small and b) integrated databases contain experimental properties of versatile molecules.

| Descriptor | Dimension | Type | |
|---|---|---|---|
| FP(Avalon) | 512 | Binary | Avalon fingerprint. |
| Desc(2D) | 200 | Continuous | Basic molecular descriptors generated by RDKit.[a] Ignores 3D structures. |
| Desc(3D) | 629 | Continuous | Descriptors related to 3D molecular strucrures, generated by DRKit.[b] |
| HSPiP | 56 | Continuous | Empirical molecular properties and descriptors calculated by HSPiP package. |
| PM7 | 13 | Continuous | Molecular properties calculated by PM7 method. |
| Neural(bp)[c] | 32 | Continuous | Neural descriptors generated by a graph neural network pretrained with the small molecule dataset to predict boiling temperature. |
| Neural(mp)[c] | 32 | Continuous | Neural descriptors for melting temperature. |
| Neural(density) | 32 | Continuous | Neural descriptors for density. |
| Neural(viscosity) | 32 | Continuous | Neural descriptors for viscosity. |

a) Available as Descriptors.descList class in RDKit.
b) Available as CalcAUTOCORR3D, CalcMORSE, CalcRDF, and CalcWHIM classes in RDKit.
c) "bp" and "mp" stands for boiling and melting temperatures, respectively.

**Figure S2.** Explanations of descriptors used in this study.

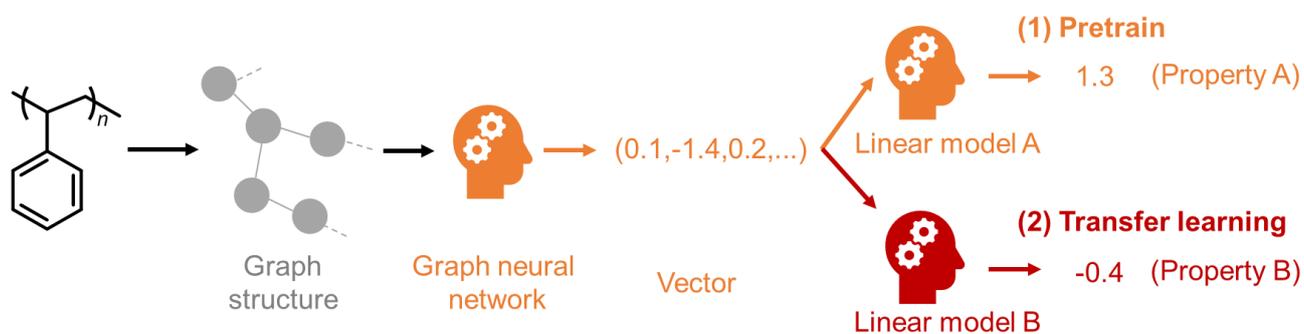

**Figure S3.** Scheme for transfer learning using graph neural networks.

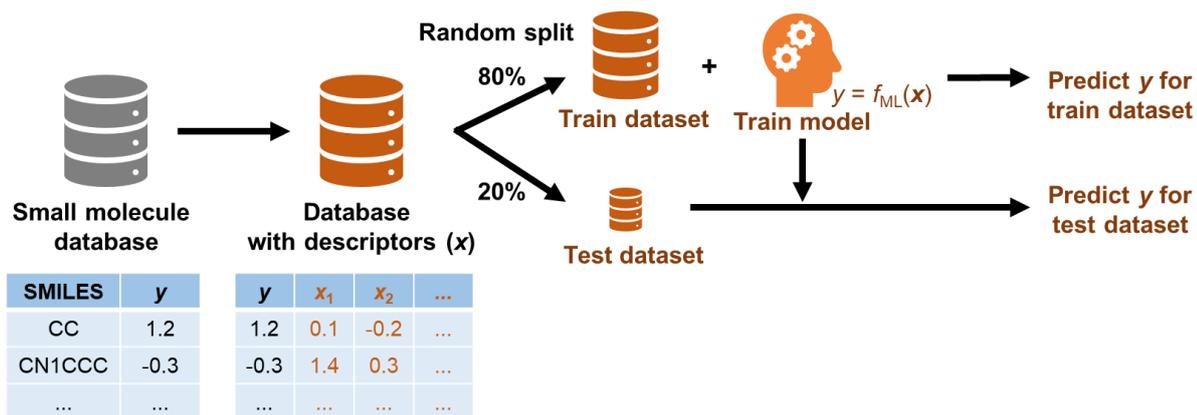

**Figure S4.** Scheme to predict molecular properties ($y$) from their descriptors ($x$).

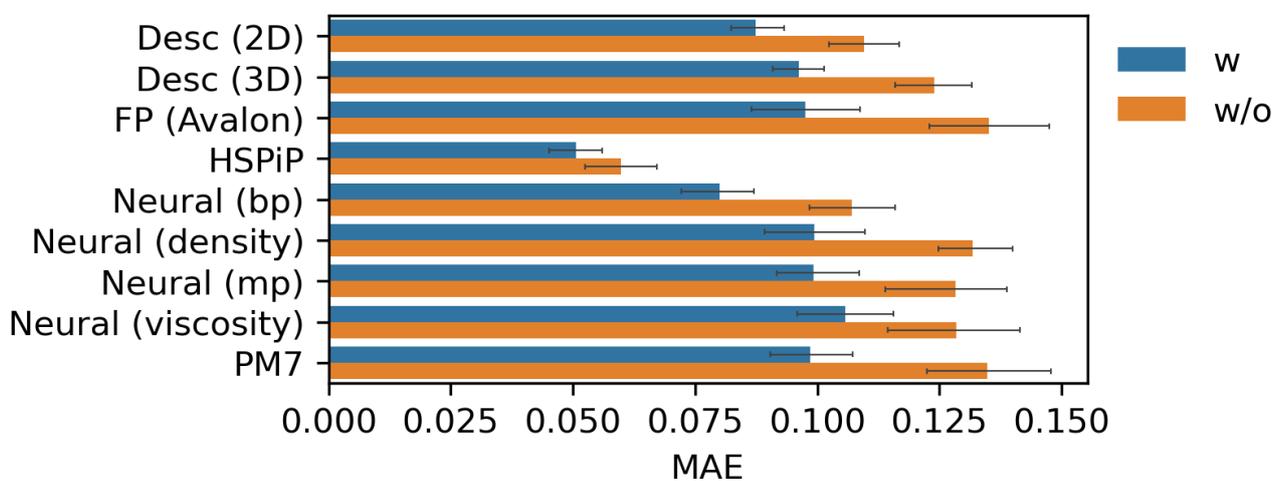

a)

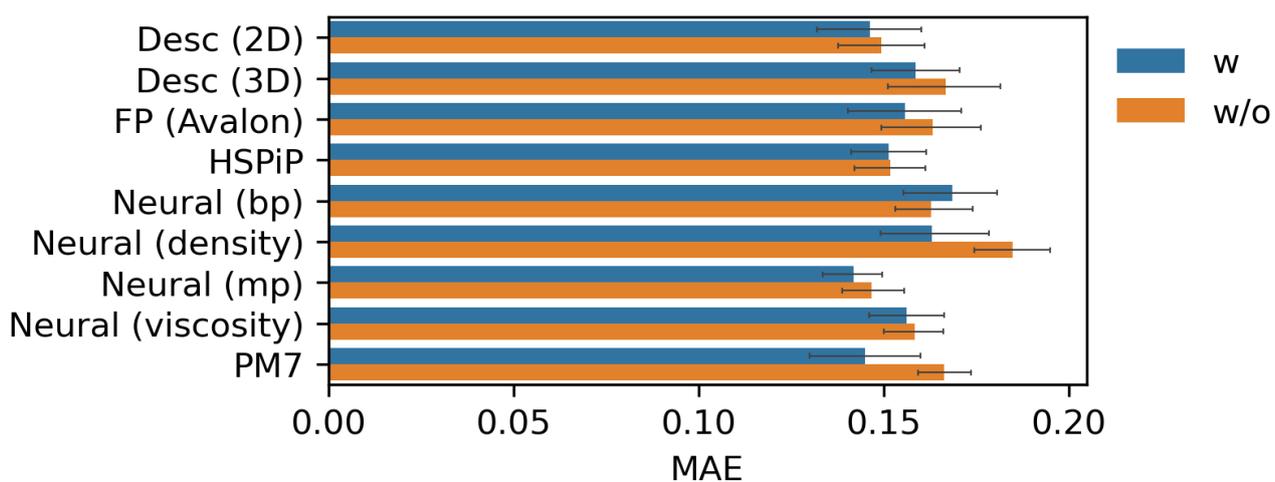

b)

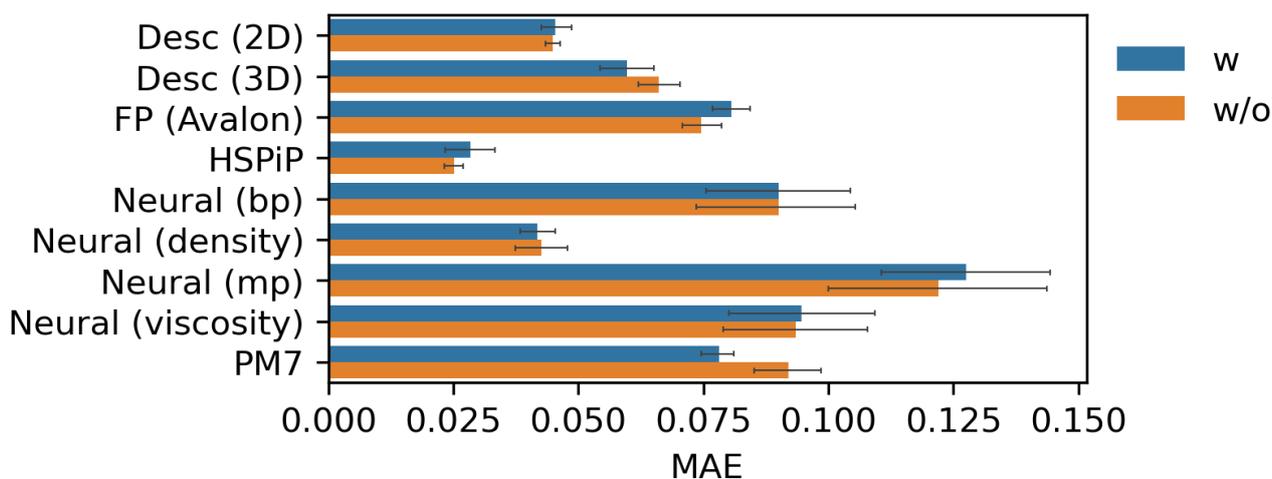

c)

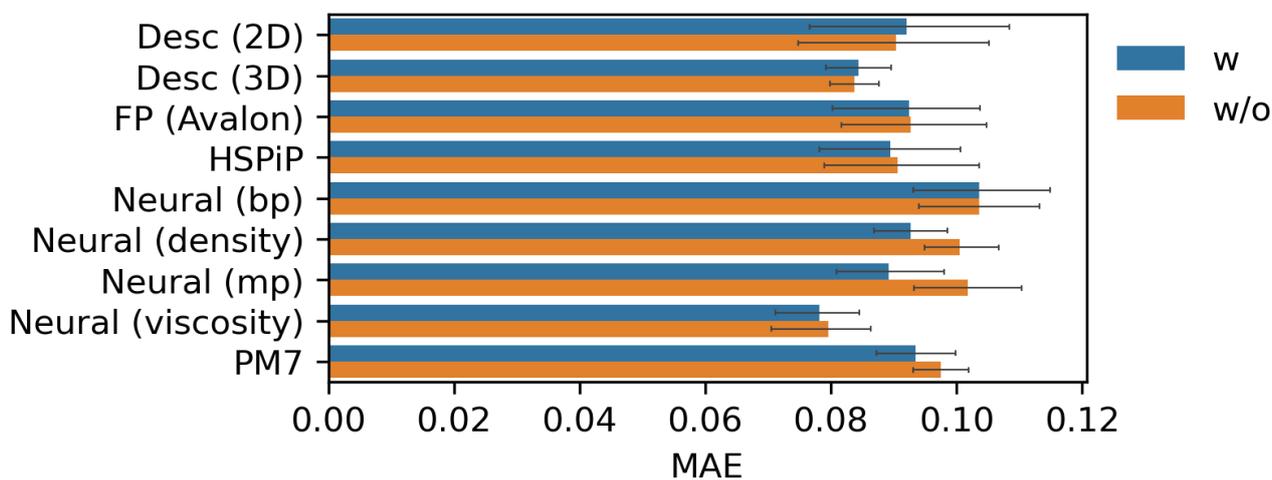

d)

**Figure S5.** Prediction results of a) boiling temperature, b) melting temperature, c) density, and d) viscosity. Average values of 5-hold cross-validation for test datasets are shown. Error bars represent standard errors. The legend "w" indicates that experimental chemical properties (i.e., boiling temperature, melting temperature, density, and viscosity) were included in $x$ in addition to original molecular descriptors, and "w/o" indicates not. XGB regressor was used as a predictor.

S7

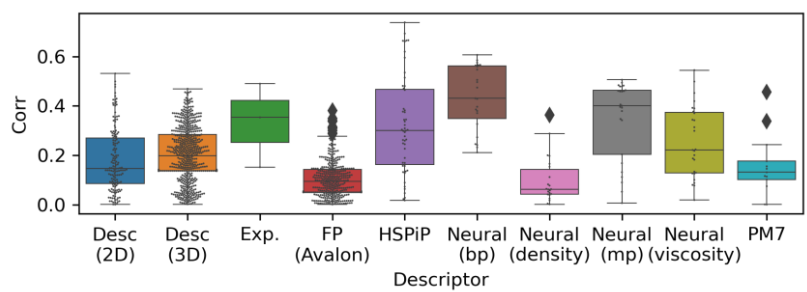

a)

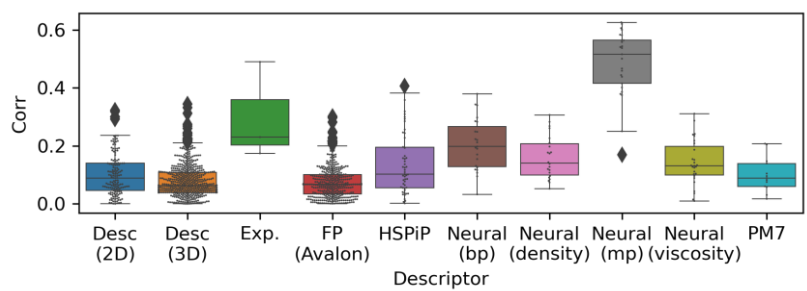

b)

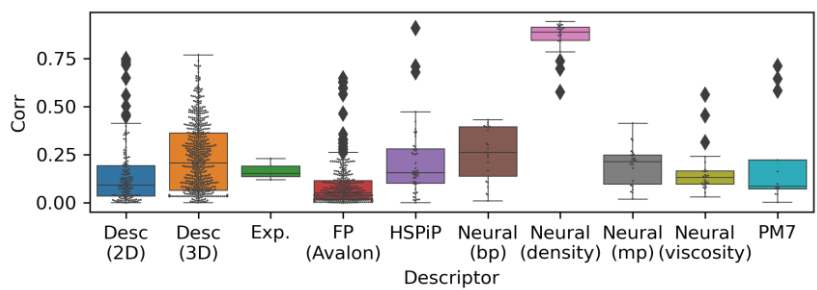

c)

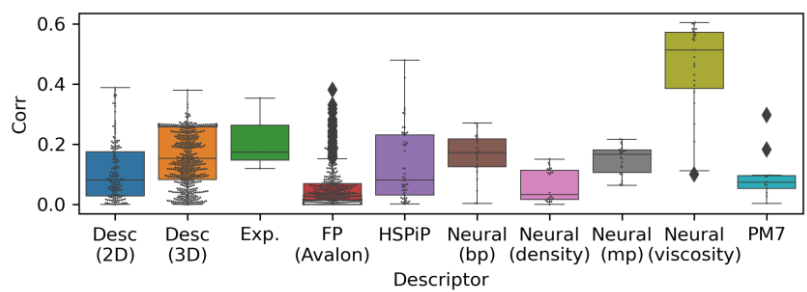

d)

**Figure S6.** Distribution of absolute correlation coefficients of molecular descriptors against a) boiling temperature, b) melting temperature, c) density, and d) viscosity. Exp, bp, and mp represent experimental chemical properties, boiling temperature, and melting temperature.

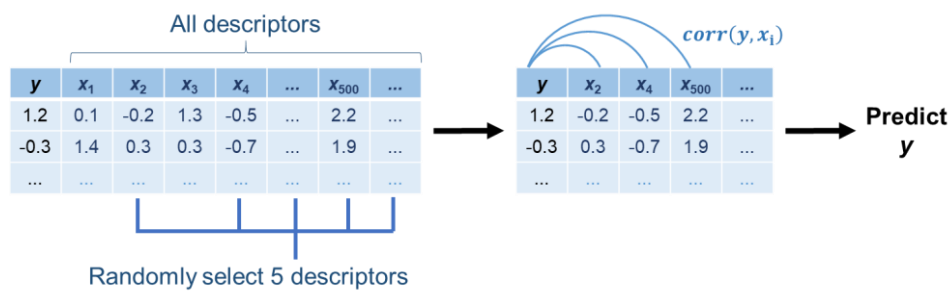

a)

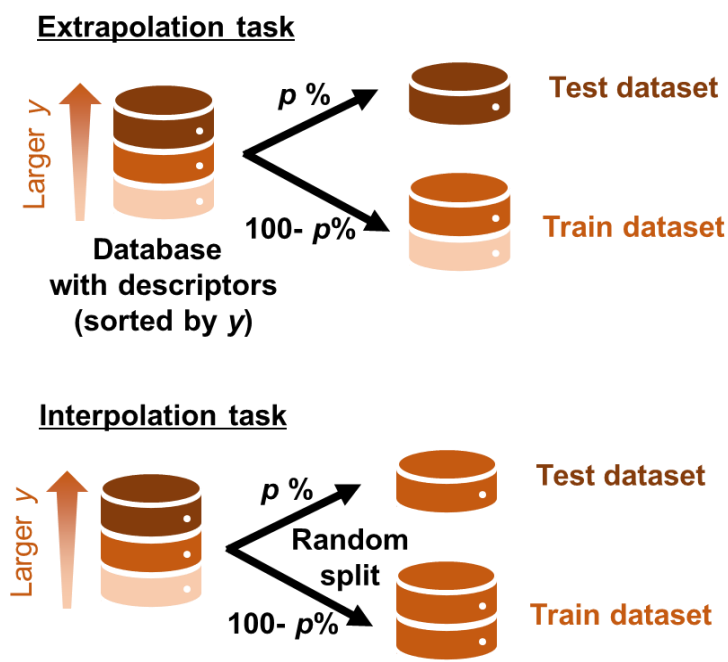

b)

**Figure S7.** a) Scheme to predict *y* from randomly selected 5 descriptors. The function corr($y,x_i$) indicates the absolute value of the correlation coefficient between y and $x_i$. Averages of corr($y,x_i$) for the selected 5 descriptors are used as the *x*-axis in **Figure S8**. b) Comparison of the extrapolation and interpolation tasks. For the extrapolation, top *p*% records of *y* were used for a test dataset ($p = 20$).

S9

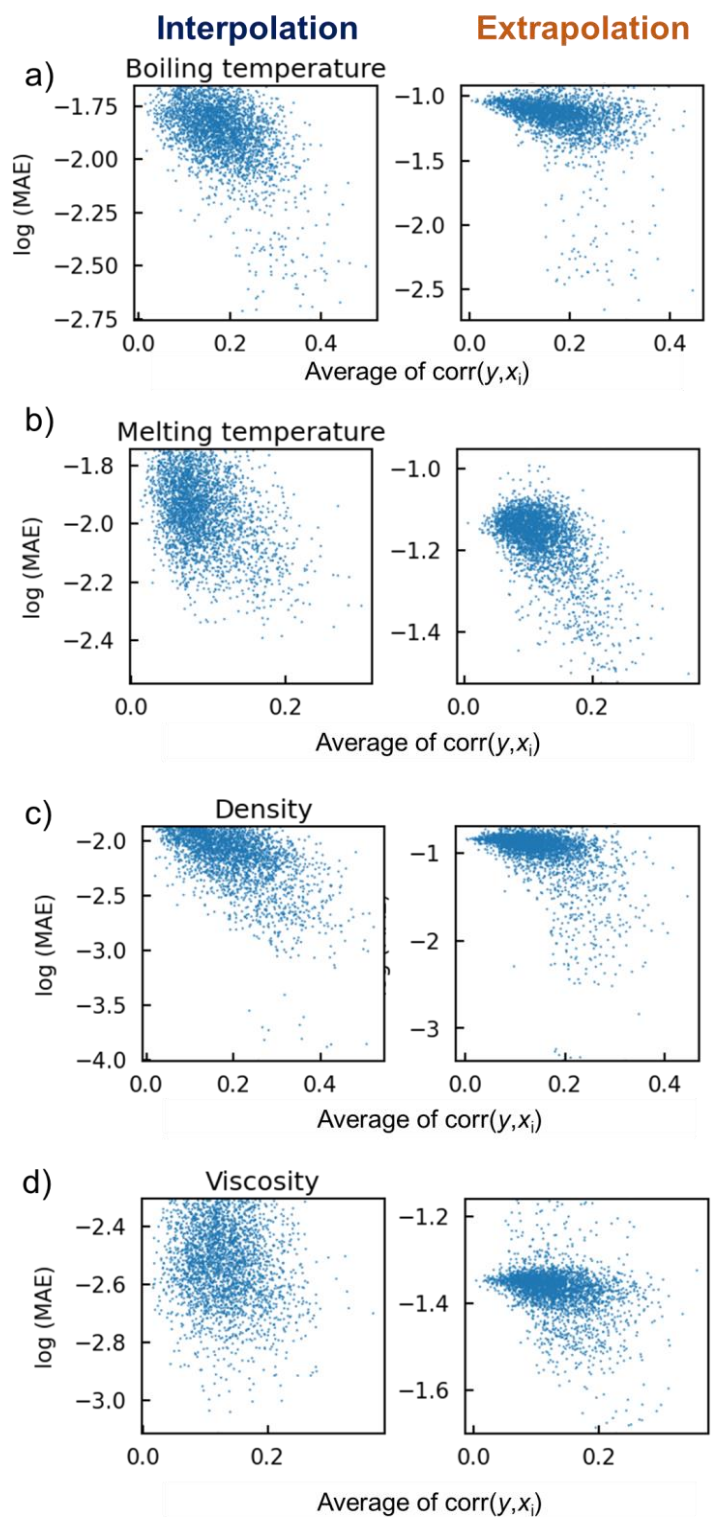

**Figure S8.** Relationship between MAE and average of corr($y, x_i$) for a) boiling temperature, b) melting temperature, c) density, and d) viscosity prediction. Datasets were prepared according to a rule shown in **Figure S7**. Huber and XGB regressors were used for interpolation and extrapolation tasks, respectively.

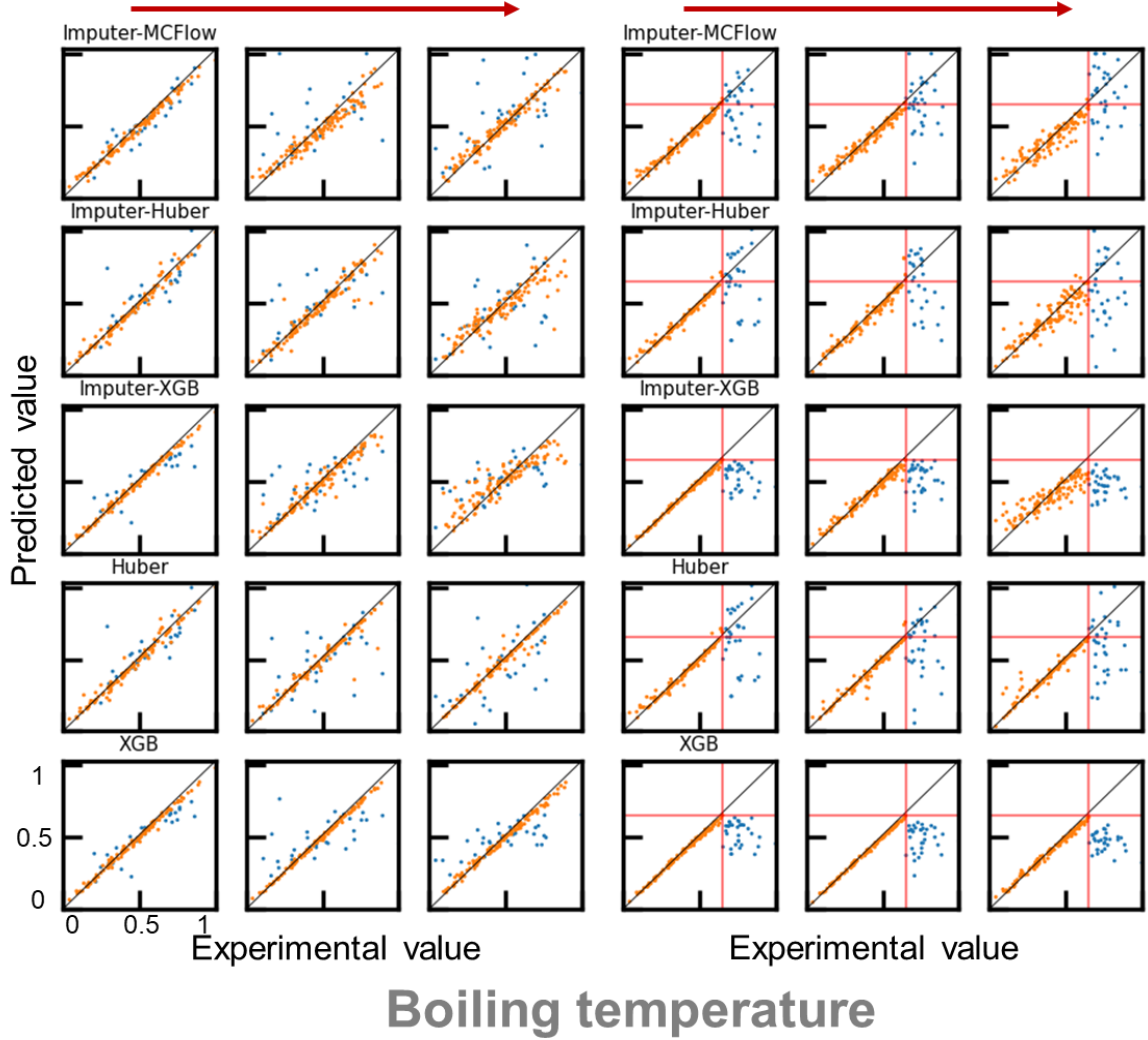

a)

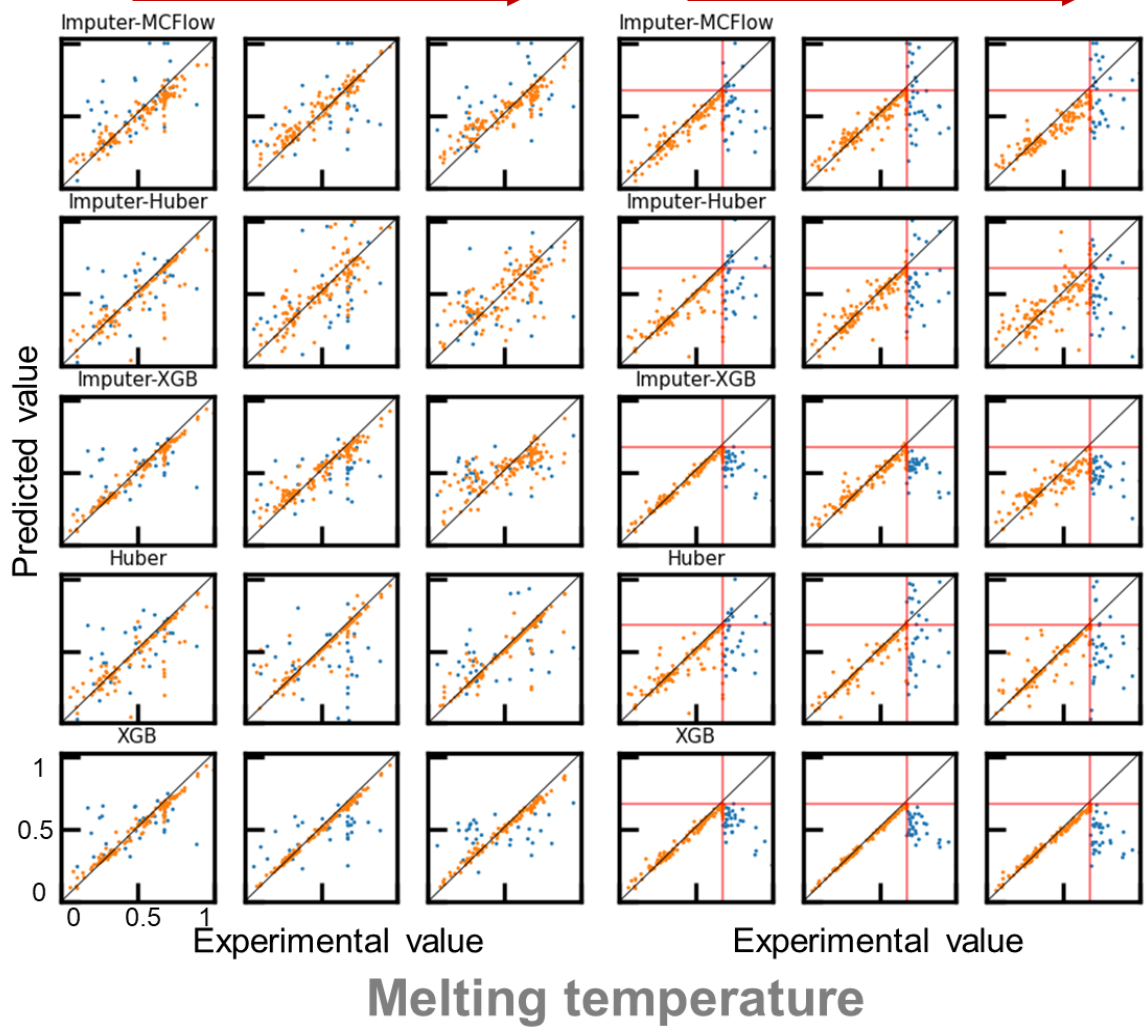

b)

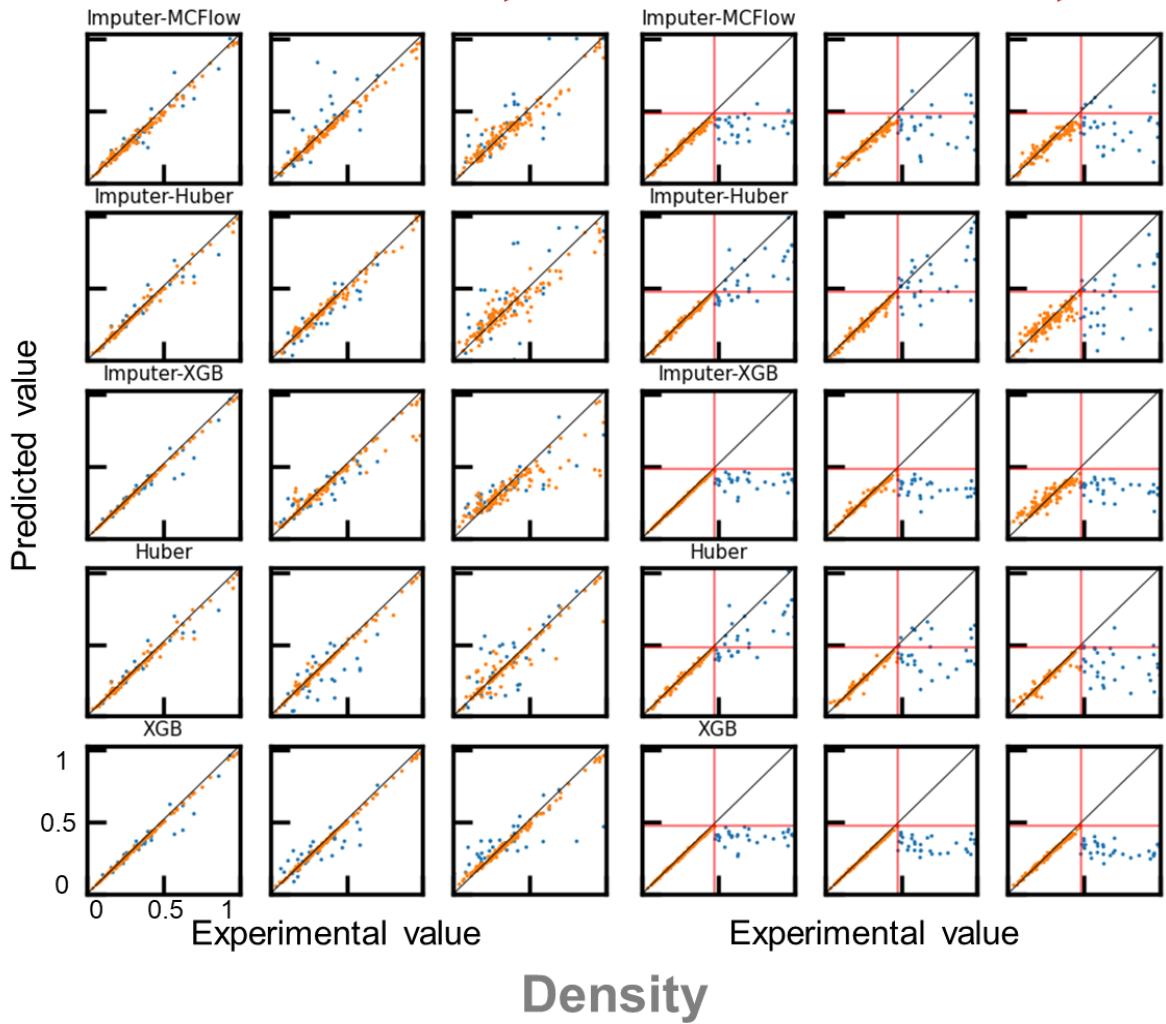

c)

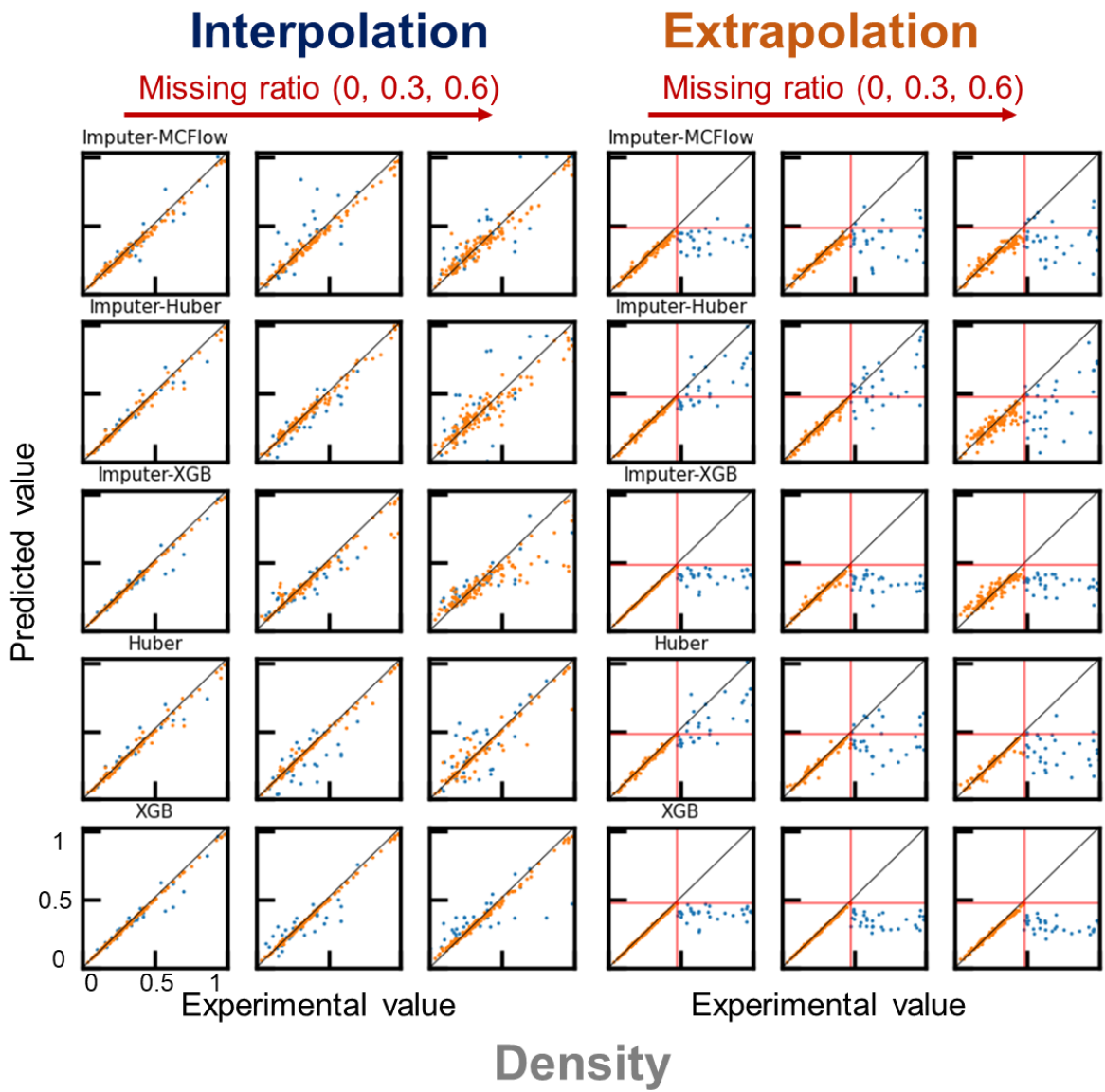

d)

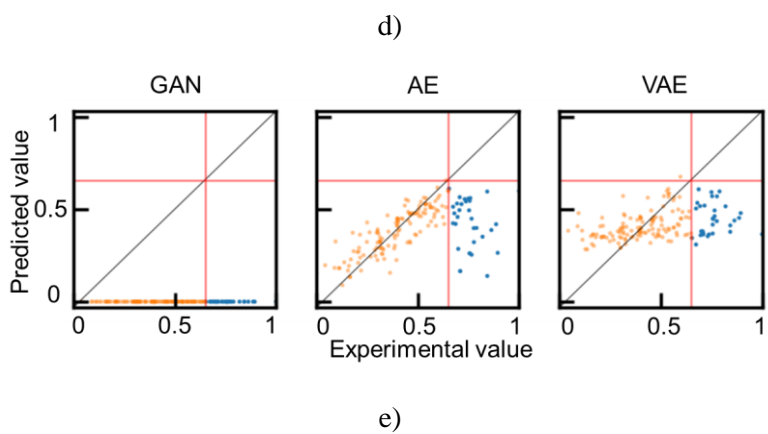

e)

**Figure S9.** Prediction results of a) boiling temperature, b) melting temperature, c) density, and d) viscosity, using different algorithms. Orange and blue plots represent train and test datasets, respectively. Before regression, $p$ (= 0, 30, or 60)% of explanatory variables ($x$) were filled with missing values (n/a) randomly.

S14

Imputers predicted *y* by imputing the missing values. Mean values of $x_i$ were used instead of the missing values for the Huber regressor. XGB regressor could input missing values directly for prediction. e) Extrapolating prediction results by GAN, AE, and VAE. Boiling temperature was predicted without data missing.

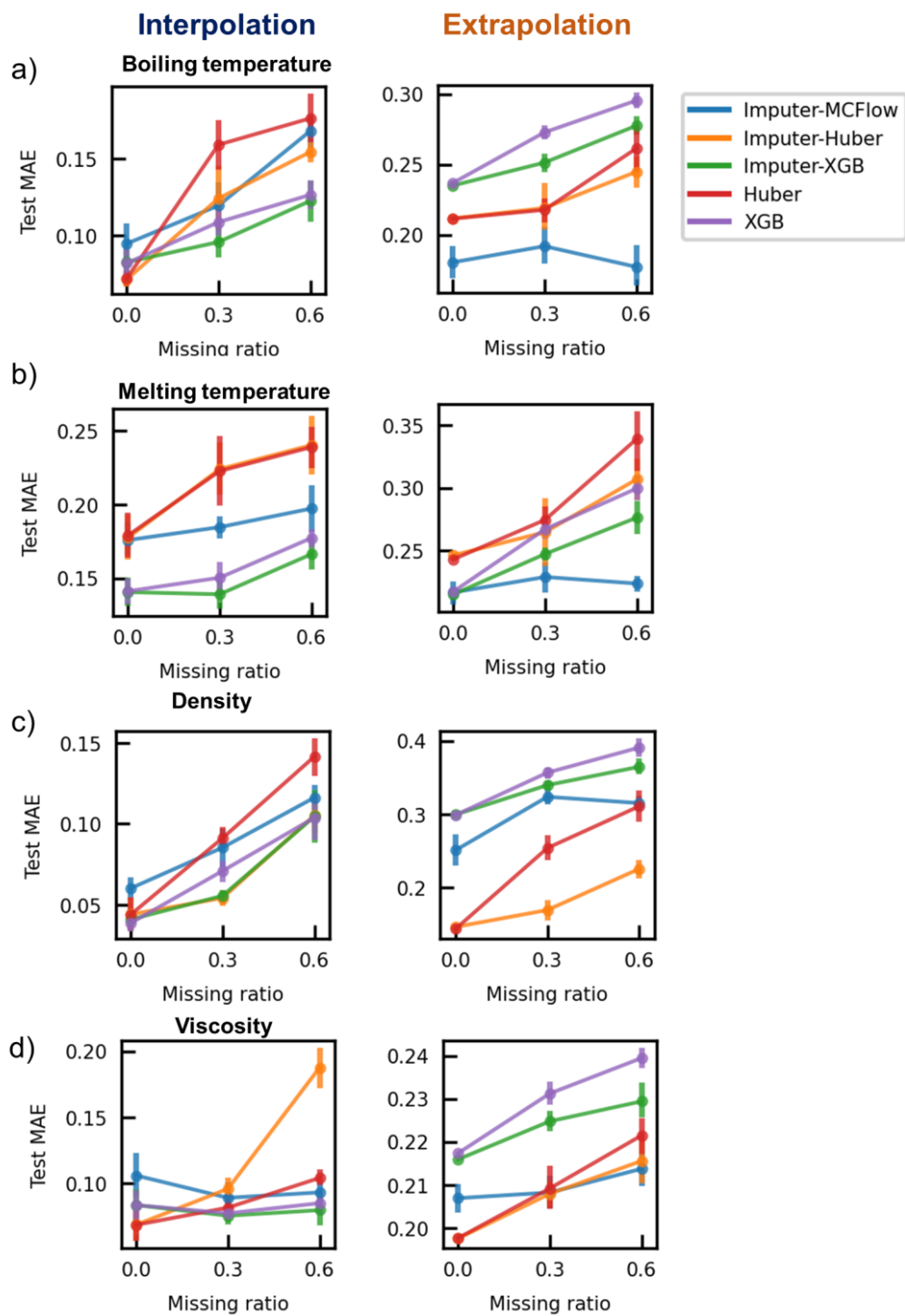

**Figure S10.** Average MAE for the prediction tasks shown in **Figure S9**. Results for the test data are presented. Dataset preparation and prediction were repeated 5 times. Error bars show standard errors.

S15

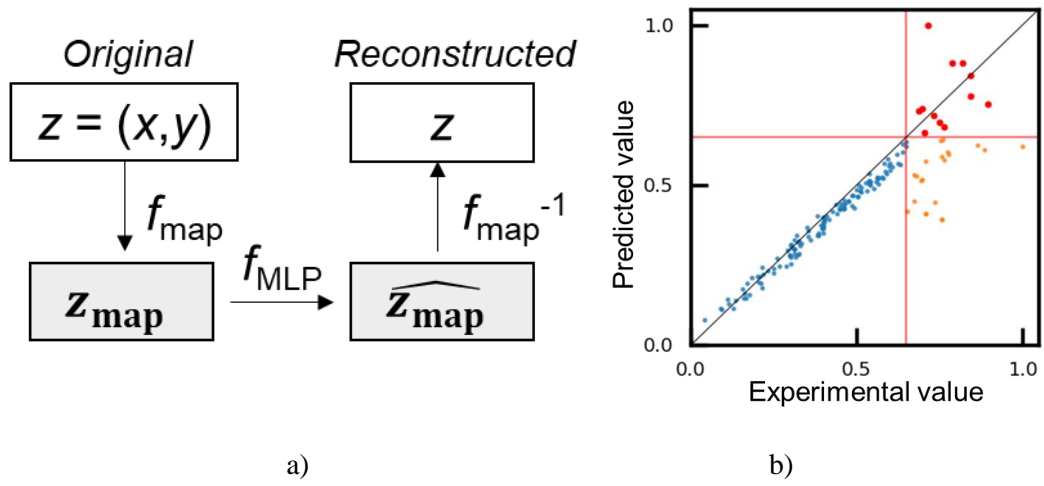

a)

b)

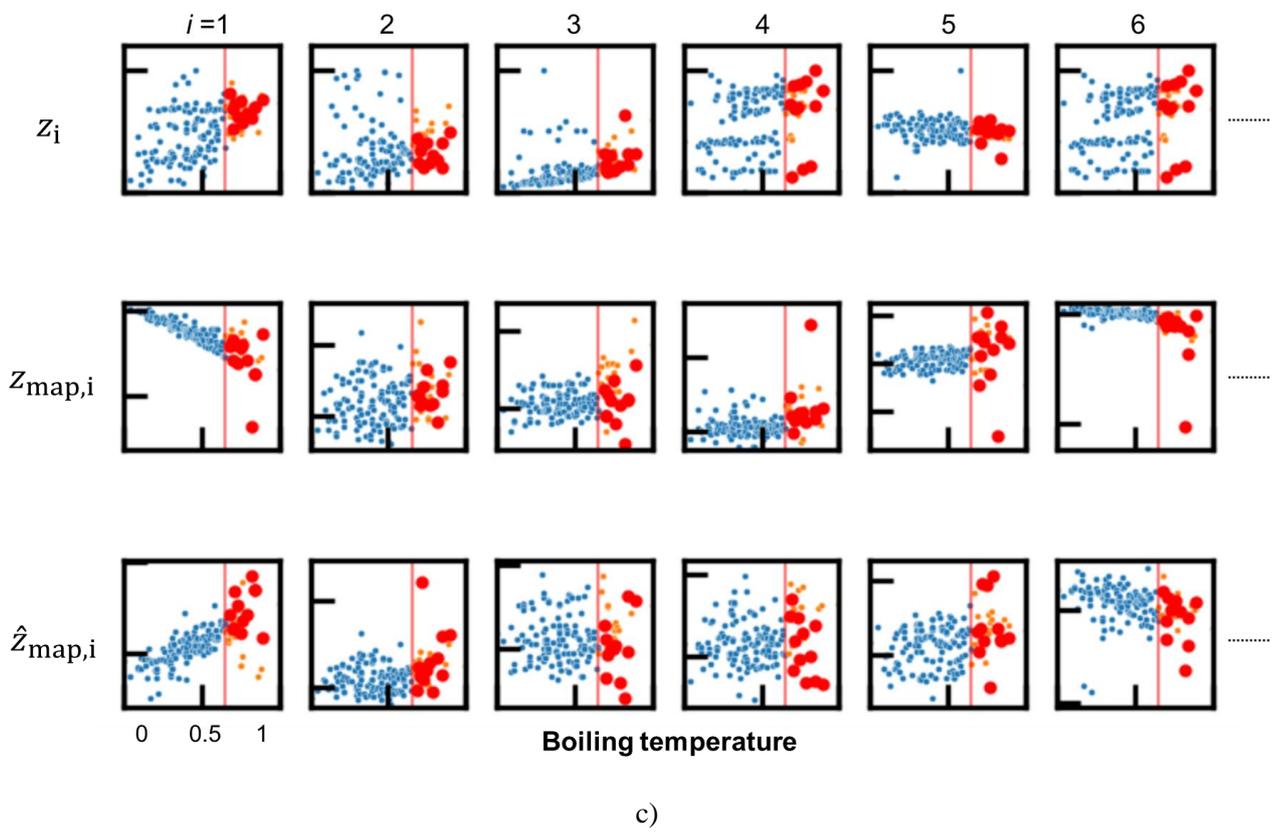

c)

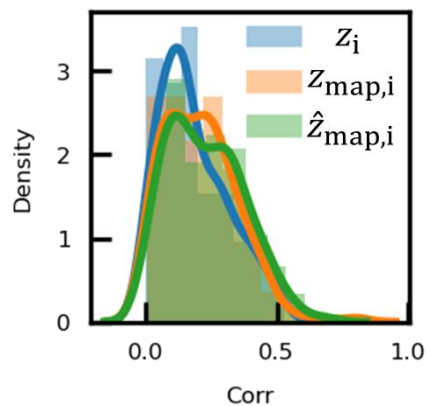

d)

**Figure S11.** a) Scheme of MCFlow. Original data ($x,y$) was first converted into $z_{map}$ having the same dimension as the original by $f_{map}$. Then, it was converted into $\hat{z}_{map}$ by a multilayer perceptron function $f_{MLP}$, and into the original space ($x,y$) by $f_{map}^{-1}$. b) Prediction result of boiling temperature (same task as **Figure S9** without data missing) by the model. c) Representative $z, z_{map}$, and $\hat{z}_{map}$ as a function of boiling temperature ($y$). d) Distribution of absolute correlation coefficients between the variables and $y$. If predicted values exceeded that of the maximum of trained $y$, they were marked by red. The model was trained to predict boiling temperature from the small compound database without data missing (i.e., the same task as **Figure S10a**).

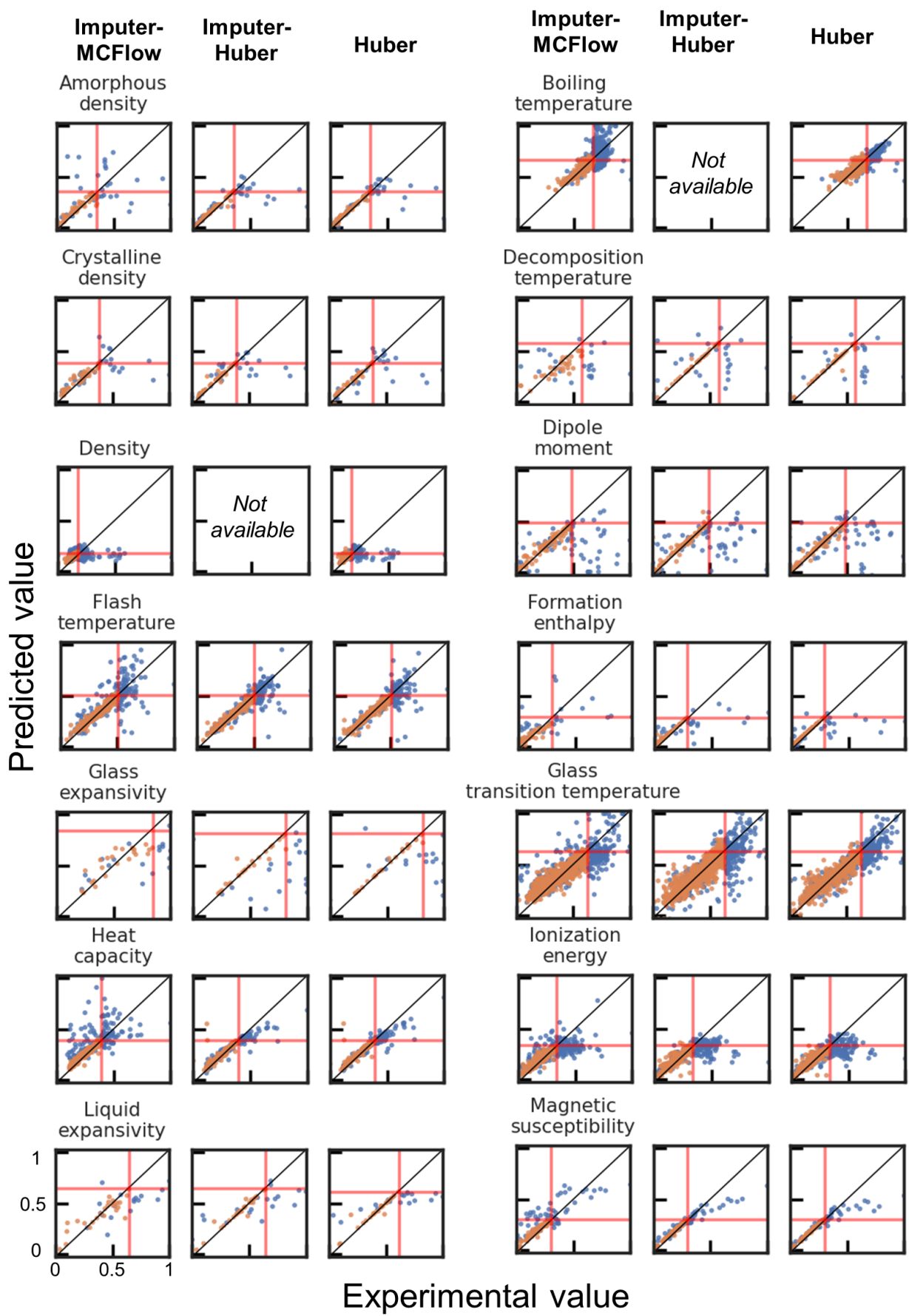

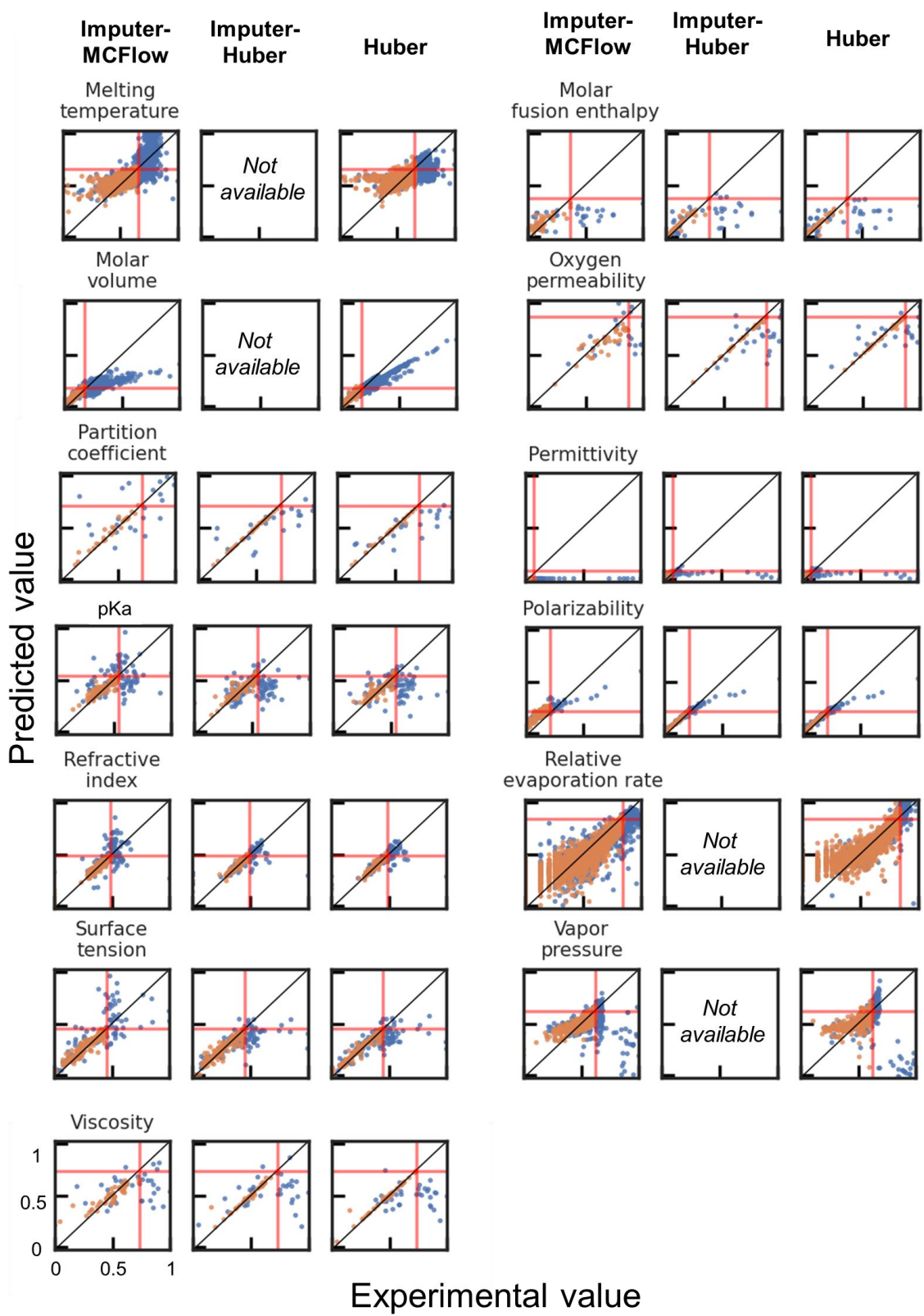

**Figure S12.** Prediction of various parameters recorded in the integrated database. Orange and blue plots represent the randomly selected 80% train and 20% test data, respectively. Some results by Imputer-Huber are not shown because of the high calculation cost of the model (e.g., the calculation will take more than 100 hours with properties containing over 10000 records).

S19

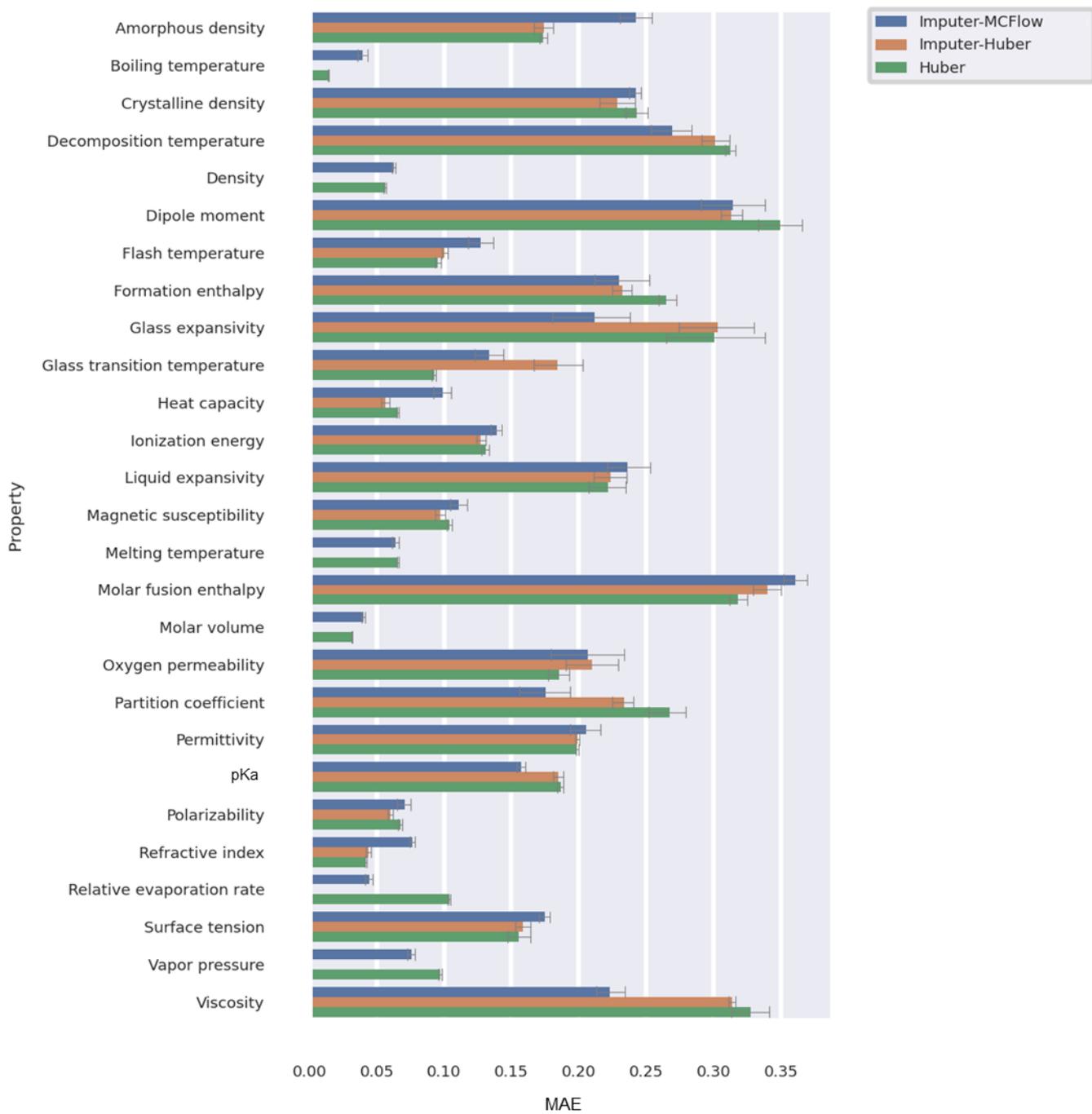

a)

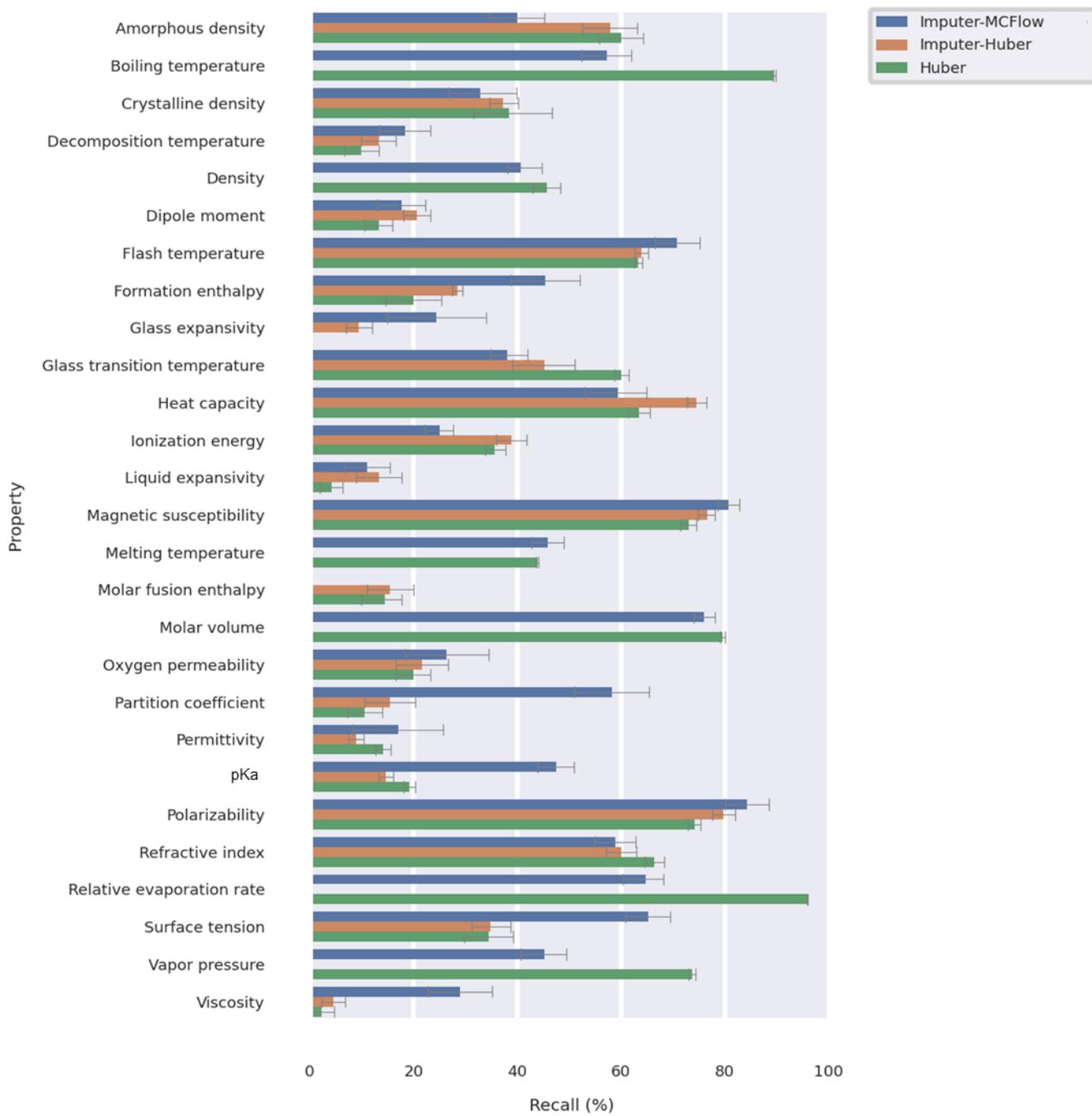

b)

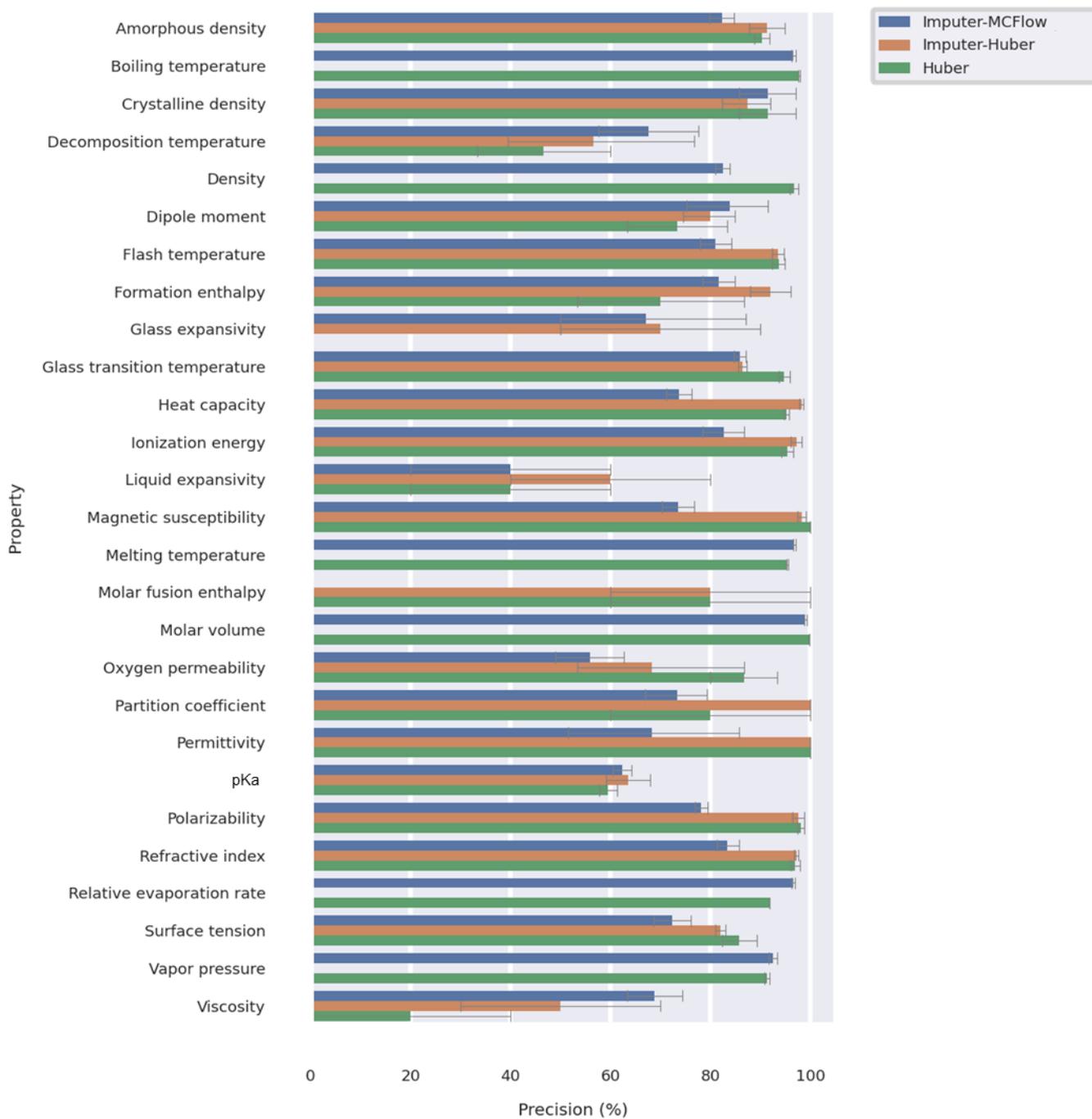

c)

**Figure S13.** Prediction performances for the test datasets of the integrated database. a) MAE for the extrapolating regions, b) recall, and c) precision. Average values for the 5-times random data splitting and regressions are shown. Error bars indicate standard errors.

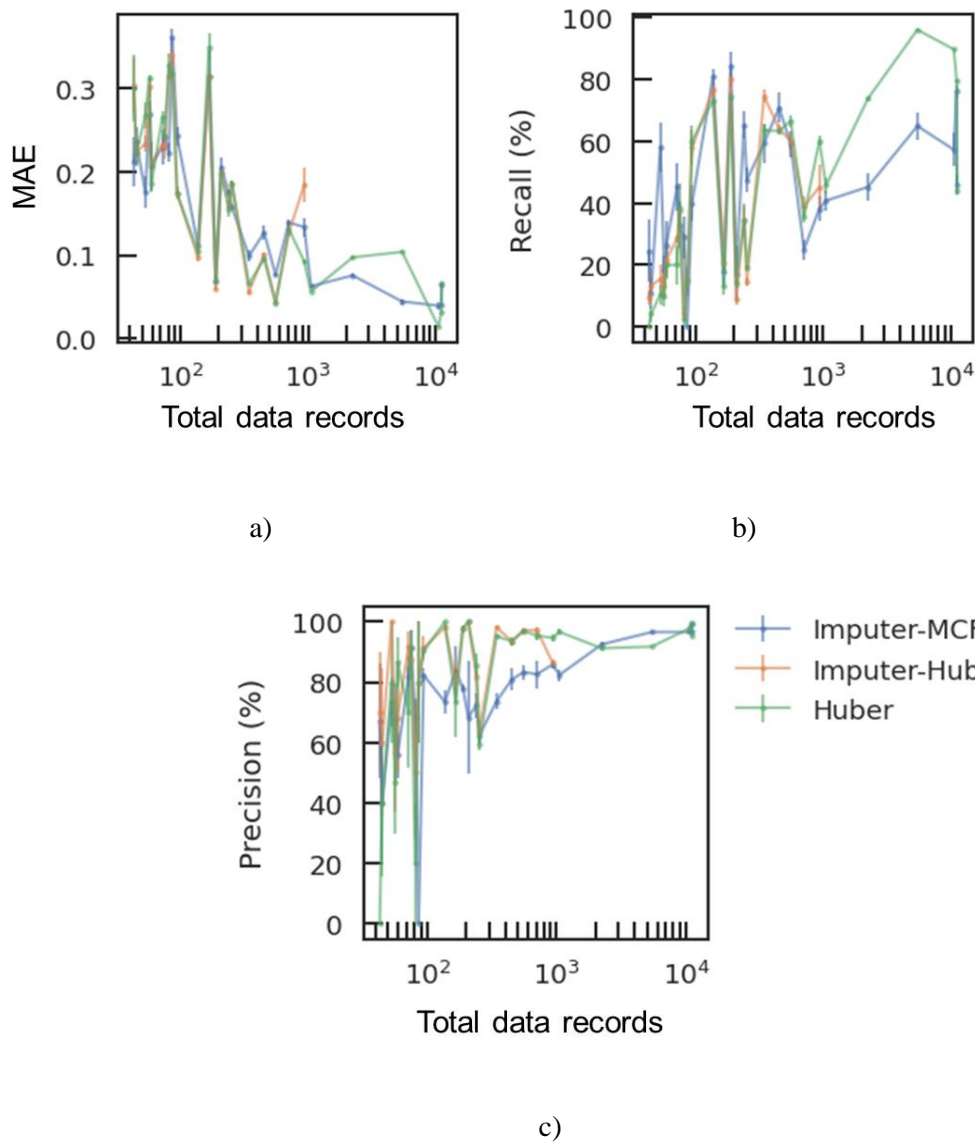

**Figure S14.** Prediction results of **Figure S13** as a function of total data records of target properties. a) MAE, b) recall, and c) precision. Error bars indicate standard errors.

**References**

(1) Hatakeyama-Sato, K.; Oyaizu, K., Integrating multiple materials science projects in a single neural network. *Commun. Mater.* **2020,** *1*, article number: 49.
(2) Stekhoven, D. J.; Buhlmann, P., MissForest--non-parametric missing value imputation for mixed-type data. *Bioinformatics* **2012,** *28*, 112-8.
(3) Richardson, T.; Wu, W.; Lin, L.; Xu, B.; Bernal, E., MCFlow: Monte Carlo Flow Models for Data Imputation. **2020,** *arXiv:2003.12628*.
(4) Yoon, J.; Jordon, J.; Schaar, M.; Xu, B.; Bernal, E., GAIN: Missing Data Imputation using Generative Adversarial Nets. **2018,** *arXiv:1806.02920*.